\documentclass[prd,reprint,preprintnumbers,amsmath,amssymb,nofootinbib,epsfig,bmamsfonts,yfonts,superscriptaddress]{revtex4-1}
\usepackage{graphicx}
\usepackage{float}
\usepackage{amsmath}
\usepackage[colorlinks=true,linkcolor=blue,citecolor=blue, urlcolor=blue]{hyperref} 
\usepackage{color}
\usepackage{soul}
\relax
\def\Fig#1{Fig.~\ref{#1}}
\def\Eq#1{Eq.~\eqref{#1}}

\def\Figs#1{Figs.~\ref{#1}}
\def\app#1{Appendix~\ref{#1}}
\def\Sec#1{Sec.~\ref{#1}}

\begin{document}

\preprint{CERN-TH-2020-127}
\title{Predicting parton energy loss in small collision systems}
\author{Alexander Huss}
\email{alexander.huss@cern.ch}
\affiliation{Theoretical Physics Department, CERN, CH-1211 Gen\`eve 23, Switzerland}
\author{Aleksi Kurkela}
 \email{a.k@cern.ch}
\affiliation{Theoretical Physics Department, CERN, CH-1211 Gen\`eve 23, Switzerland}
 \affiliation{Faculty of Science and Technology, University of Stavanger,  4036 Stavanger, Norway}
\author{Aleksas Mazeliauskas}
\email{aleksas.mazeliauskas@cern.ch}
\affiliation{Theoretical Physics Department, CERN, CH-1211 Gen\`eve 23, Switzerland}
\author{Risto Paatelainen}
\email{risto.sakari.paatelainen@cern.ch}
\affiliation{Theoretical Physics Department, CERN, CH-1211 Gen\`eve 23, Switzerland}
\author{Wilke van der Schee}
\email{wilke.van.der.schee@cern.ch}
\affiliation{Theoretical Physics Department, CERN, CH-1211 Gen\`eve 23, Switzerland}
\author{Urs Achim Wiedemann}
 \email{urs.wiedemann@cern.ch}
\affiliation{Theoretical Physics Department, CERN, CH-1211 Gen\`eve 23, Switzerland}

\begin{abstract} 
Medium induced parton energy loss is not conclusively established either in very peripheral heavy-ion collisions or in proton-ion collisions.
However, the standard interpretation of azimuthal momentum anisotropies  in theses systems implies some partonic rescattering.
The upcoming light-ion runs at the Large Hadron Collider (LHC) provide a unique opportunity to search for parton energy loss in different systems of similar size.
Here, we make predictions for the expected parton energy loss signal in the charged hadron spectra in a system size scan at LHC. 
We test a large set of model assumptions against the transverse momentum and centrality dependence of the charged hadron nuclear modification factor in lead-lead and xenon-xenon collisions at the LHC. 
We then attempt to make a model agnostic prediction for the charged hadron nuclear modification factor in oxygen-oxygen collisions. 
\end{abstract}

\maketitle
\section{Introduction}
\label{sec1}
The observed factor $5$ suppression of the charged hadron nuclear modification factor $R_\text{AA}^{h}$ in 
central $\sqrt{s_{NN}} = 130$ GeV Au-Au collisions at the Relativistic Heavy Ion Collider (RHIC) marked the start 
of experimental energy loss studies two decades ago~\cite{Adcox:2001jp,Adler:2002xw}. Pb+Pb collision data from the Large Haddron Collider (LHC) showed that this quenching 
increases mildly with center-of-mass energy, and that nuclear modifications remain visible in hadron spectra up to the transverse momentum $p_\perp \approx {\cal O}\left(100\, {\rm GeV}\right)$ \cite{Aamodt:2010jd,Abelev:2012hxa,CMS:2012aa,Aad:2015wga}. 
An important early finding at RHIC  was that (within experimental uncertainties) quenching disappears in $d$+Au collisions
 where no dense medium was expected to interact with high-$p_\perp$
partons in the final state~\cite{Adler:2003ii,Adams:2003im,Arsene:2003yk}. This finding was later corroborated at LHC where 
quenching is absent in TeV-scale $p$Pb collisions~\cite{ALICE:2012mj,Abelev:2014dsa,Khachatryan:2015xaa,Aad:2016zif}. 

A reassessment of the conclusions drawn from these
data in small systems may be needed in the light of the recent LHC discovery of strong collectivity (``flow'') in soft
multihadron correlations~\cite{Abelev:2014mda,Khachatryan:2015waa,Sirunyan:2017uyl,Aaboud:2017acw}, 
and its confirmation in the subsequent analysis of small collision systems at RHIC~\cite{Adare:2014keg,Adamczyk:2015xjc}. 
According to the standard phenomenological interpretation, $v_n$ measurements indicate significant final state interactions between colored degrees of freedom in small collision systems.
This raises the questions of why high-$p_\perp$ energy loss effects have escaped so far experimental detection in small systems and how such effects could be revealed in future experiments. To address this, our 
paper develops and documents parton energy loss models that 
extend to the smallest hadronic collision systems.

Parton energy loss in the QCD medium was predicted in the pioneering works of Bjorken~\cite{Bjorken:1982tu} and of Gyulassy, Pluemer and 
Wang~\cite{Wang:1994fx,Gyulassy:1993hr}. It was given a first QCD-based treatment by Baier, Dokshitzer, Mueller, 
Peign\'e and Schifff (BDMPS)~\cite{Baier:1996kr,Baier:1996sk}, and by Zakharov (Z)~\cite{Zakharov:1996fv,Zakharov:1997uu}, 
with later refinements by others~\cite{Wiedemann:2000za,Gyulassy:2000er,Wang:2001ifa}. 
These works calculate, 
for an arbitrary number of interactions with the
medium, the non-Abelian Landau-Pomeranchuk-Migdal (LPM) effect that underlies medium induced parton splitting. 
The same LPM effect was found independently by
Arnold, Moore and Yaffe (AMY) when developing an effective kinetic transport formulation of hard degrees of freedom
in QCD finite temperature field theory~\cite{Arnold:2001ba,Arnold:2001ms}. 
Spurred by the measurement
of quenched jets (as opposed to quenched high-$p_\perp$ hadrons) at the LHC, much subsequent theoretical 
work aimed at extending the BDMPS-Z formalism to multiparton final states, either by encoding jet quenching in Monte Carlo 
simulations~\cite{Zapp:2008gi,Zapp:2012ak,Zapp:2013vla,Armesto:2009fj,Schenke:2009gb,Caucal:2018ofz,Putschke:2019yrg} 
or by extending the BDMPS-Z formalism to higher order in $\alpha_s$ and thus to higher number
of medium induced gluons in the final state~\cite{Arnold:2015qya,Arnold:2016kek,CasalderreySolana:2012ef,Caucal:2018dla}. 

In the present paper we focus on modeling the suppression of high-momentum hadron spectra. Our starting point is a particularly
clean and simple reformulation of the BDMPS-Z formalism due to Arnold~\cite{Arnold:2008iy} from which we determine the
probability distribution of parton energy loss (``quenching weight'') and the resulting hadron nuclear modification factor 
following Ref.~\cite{Baier:2001yt}. 
There have been several model comparisons to quenched hadron spectra
with the systematic study of the centrality dependence of the nuclear modification factor~\cite{Chien:2015vja,Bianchi:2017wpt,Andres:2016iys,Noronha-Hostler:2016eow,Casalderrey-Solana:2014bpa, Djordjevic:2015hra, Zigic:2018smz,Andres:2019eus}.
These works focus on the centrality in PbPb and XeXe (AuAu)
collisions at the LHC (at RHIC). Our aim is to validate an energy
loss model on this centrality dependence and 
to use it for predicting nuclear modification factors in the foreseen
TeV-scale minimum bias collisions of lighter nuclei, i.e., in oxygen-oxygen (OO) and argon-argon (ArAr) collisions~\cite{Citron:2018lsq}.
 For the heavy quark nuclear modification in small systems, a similar approach has been followed in Ref.~\cite{Katz:2019qwv}.
For the nuclear modification of jets, 
several studies of $p$Pb at the LHC~\cite{Zhang:2013oca,Park:2016jap,Sarkar:2018esp,Pierog:2013ria} (see also Ref.~\cite{Tywoniuk:2014hta}) arrived at quenching effects that are larger than the current bounds set by experiments. A community-wide study of future physics
opportunities for high-density QCD at the LHC~\cite{Citron:2018lsq} asked for further modeling efforts,
noting that current Monte Carlo
models of parton energy loss~\cite{Zapp:2013vla} may somewhat
over predict medium effects in argon and xenon collisions.

Theoretical uncertainties in applying the BDMPS-Z formalism to quenched hadron spectra have been analyzed in a community-wide study~\cite{Armesto:2011ht}, and they have been included in 
subsequent extractions of the jet transport coefficient $\hat q$ from data~\cite{Burke:2013yra,Andres:2016iys}. 
In  addition, there are known event selection and geometry biases that in peripheral AA collisions complicate the model 
comparison of nuclear modification factors~\cite{Morsch:2017brb}. 
One qualitative conclusion of the present paper will be that an energy loss model based on the BDMPS-Z formalism and consistent with experimental data in PbPb and XeXe collisions can result in sufficiently small nuclear modifications in OO collisions that a high accuracy baseline is needed to detect medium induced energy loss. In our companion paper \cite{Huss:2020dwe} we show that this is indeed possible.

In Sec.~\ref{sec2}, we shall provide a 
description of different building blocks of a parton energy loss model. 
We also comment on the system size dependence of theoretical uncertainties.  
Section~\ref{sec3} presents our results on momentum  and system size dependence of the charged hadron nuclear modification factor. Because our simplified model does not take into account all the
details of modeling soft QCD medium evolution in heavy ion collisions, we vary various model assumptions to test the robustness of our predictions. Although it is not the main focus of our paper,
we also checked the model predictions for high-momentum hadron $v_2(p_\perp)$.
Our conclusions are given in Sec.~\ref{conclusions}.

\section{Simple parton energy loss model} 
\label{sec2}

Most formulations of parton energy loss for single inclusive hadron production start from the framework of collinearly factorized perturbative QCD. In this framework, a generic hadronic cross section can be schematically written as
\begin{equation}
\label{schematicv2}
\sigma^h = \hbox{PDFs} \otimes \sigma^{\text{vac}}_{\rm g/q}  \otimes \hbox{FFs},
\end{equation}
where the perturbatively computable hard partonic [gluon $(g)$ and quark $(q)$] cross sections $\sigma_{g/q}^{\rm vac}$ are convoluted with the universal process-independent parton distribution functions (PDFs) that describe the parton content of the hadrons and with the hadronic fragmentation functions (FFs). This starting point provides a systematically improvable baseline for calculating the spectra in the absence of medium effects. 

Nuclear effects in \Eq{schematicv2} enter in two ways. First, the parton distribution functions in ultra-relativistic colliding nuclei differ characteristically from those in free protons, and hence, the PDFs are replaced by nuclear PDFs (nPDFs)~\cite{Eskola:2016oht,Kovarik:2015cma,AbdulKhalek:2019mzd,Walt:2019slu,deFlorian:2011fp}. Second, the partons leaving
the high-momentum transfer vertex of a nucleus-nucleus collision enter a dense QCD medium that affects their parton shower. 
In the description of single inclusive hadron spectra, this 
is typically modeled 
by replacing the hard partonic \emph{vacuum} cross section by a medium-modified differential parton cross section
\begin{align}
    \frac{d \sigma_{g/q}^{\rm med}}{dyd p_\perp^2} = \int d\epsilon  P_{g/q}(\epsilon) \frac{d \sigma_{g/q}^{\rm vac}(p_\perp + \epsilon)}{dyd p_\perp^2}\, .
\end{align}
Here, $P_{g/q}(\epsilon)$ denotes the  probability for a gluon (quark) with momentum $p_\perp+\epsilon$ to lose $\epsilon$ of its transverse momentum prior to being convoluted with the fragmentation function. 

The nuclear modification of centrality averaged hadron spectra is expressed as the ratio of charged hadron cross sections in nucleus-nucleus (AA) collisions and $pp$ collisions scaled by $A^2$, where $A$ is the total number of neutrons and protons in the nucleus:
\begin{equation}
     R_\text{AA}^{h}(p_\perp, y) =\frac{1}{A^2} \frac{d\sigma^{h}_\text{AA}/dydp^2_\perp}{d\sigma^{h}_{pp}/dydp^2_\perp}.
     \label{RAA}
\end{equation}
The hadron nuclear modification factor is the main deliverable of our simple energy loss model. 
We work at mid-rapidity $|y|<1$ and  drop  the explicit $y$-dependence in the following.

In the subsequent sections we describe in detail different model assumptions entering  $P_{g/q}(\epsilon)$ and how  \Eq{RAA} is computed in the presence of medium modifications.

\subsection{Medium induced gluon radiation}

Inelastic processes provide the most efficient mechanism for degrading the energy of high-momentum partons. In models of radiative parton energy loss, these are described by calculating the medium induced  
gluon emission rate 
$d I^{g/q}_\text{med}/{d \omega}$~\cite{Wang:1994fx,Gyulassy:1993hr,Baier:1996kr,Baier:1996sk,Zakharov:1996fv,Zakharov:1997uu,Wiedemann:2000za,Gyulassy:2000er,Wang:2001ifa}. 
Following Ref.~\cite{Baier:2001yt}, the probability $P_{g/q}(\epsilon)$ is given as a sum over the probability to emit $n$ medium-induced bremsstrahlung
gluons $\epsilon=\sum_{i=1}^n \omega_i$, 
\begin{eqnarray}
    P_{g/q}(\epsilon) &=& \sum_{n=0}^\infty \frac{1}{n!} \left[\prod_{i=1}^n \int_0^\infty d \omega_i 
    \frac{d I^{g/q}_\text{med}}{d \omega_i} \right] \delta(\epsilon - \sum_{i=1}^{n} \omega_i) \nonumber \\
    && \times \exp\left(-\int_0^\infty d\omega \frac{dI^{g/q}_\text{med}}{d \omega}\right).
    \label{quenchweight}
\end{eqnarray}
The factorial accounts for an arbitrary ordering of the emissions and the exponential normalizes the distribution to $\int_0^\infty d\epsilon P(\epsilon)=1$.

Here, we use for the evaluation of the medium induced
gluon emission rate a particularly clean and transparent
reformulation of the BDMPS-Z formalism due to Arnold~\cite{Arnold:2008iy}.
For a high-energy parton of species $s$ with energy $E$ moving through a medium, we write~\cite{Arnold:2008iy} 
\begin{align}
    \omega \frac{d (I^s- I^s_\text{vac})}{d \omega } \equiv \omega \frac{d I^s_\text{med}}{d \omega } = \frac{\alpha_s  }{\pi} x P_{s\rightarrow g}(x) \ln | c(0)|, \label{dI}
\end{align}
where $x$ is the momentum fraction carried by the emitted gluon, and $s=g/q$ denotes the species of the emitting parton. 
In the vacuum, this gluon emission is dictated by the Dokshitzer-Gribov-Lipatov-Altarelli-Parisi vacuum splitting function $P_{s \rightarrow g}$.  The factor $\ln | c(0)|$ determines to what extent the gluon emission rate $dI^s$ in the medium differs from that in the vacuum. The entire BDMPS-Z formalism can be reduced to the problem of determining $|c(0)|$ from the function $c(t)$, which satisfies the differential equation~\cite{Arnold:2008iy}
\begin{align}
\frac{ d^2 c}{dt^2} = - \omega_0^2(t)c(t)
\label{Arnolddiff}
\end{align}
with the boundary condition that $c(t) \rightarrow 1$ and $c'(t) \rightarrow 0$ for $t\rightarrow \infty$. Here, the complex frequency $\omega_0(t)$ is given in the small $x\ll 1$ limit by
\begin{align}
    \omega_0^2(t) = - i \frac{(1-x)C_A + x^2 C_s}{2x(1-x) E}\hat{\bar q} \approx  - i \frac{C_A}{2 \omega }\hat{\bar q}(t ,\vec x(t)) \,,
    \label{omega0}
\end{align}
where $\omega = xE$ is the energy of the radiated gluon. For small $x$ we have $xP_{s\to g}(x)\approx C_s$. 

All information about the interaction with the QCD medium enters the formalism via the quenching parameter $\hat{\bar q}$ in \Eq{omega0}.
 This parameter, multiplied by the Casimir $C_s$ of the corresponding color representation of the energetic parton,
characterizes the average transverse momentum squared  $\hat q = C_s \hat{\bar q} $ that is transferred due to soft
interactions from the QCD medium to the energetic parton per unit path length.  To leading order (LO) in the weak coupling
expansion, $ \hat{\bar q}$ is independent of the particle species. It depends in general
on the local density that the medium has at time $t$ at position $\vec x(t)$,
where $\vec x(t)$ is the trajectory of the hard parton through the medium. In this way, information about
the density of the soft QCD medium and its time evolution enters the calculation of modified high-$p_\perp$ hadron spectra.

\subsection{Background temperature parametrization}
\label{sec:geometry}

Many sophisticated hydrodynamic models exist for the evolution of the bulk QCD medium that have been validated phenomenologically
against soft physics data in central and semi-peripheral collisions. In principle, any of these models could be interfaced with the present formalism via a simple prescription that determines 
$\hat{\bar q}(t ,\vec x(t))$ from the soft bulk quantities evolved.
However, in very peripheral collisions of $90 \%$ centrality and light-ion collisions (with number of participant nucleons
$\langle N_{\rm part} \rangle \approx 10$) the assumptions about the fluid dynamic
evolution of QCD matter may become more questionable.

Without entering a detailed discussion about the system size dependence of the soft physics modeling~\cite{Adolfsson:2020dhm}, we 
employ a particularly simple setup of the QCD medium evolution
in which the system size dependence is given in terms of 
a few parameters. We will subsequently  vary the background evolution 
to gain insight into the robustness of the parton energy loss signal.
For background temperature evolution $T(\tau, \vec x_\perp)$ we use a one-parameter (opacity $\hat{\gamma}$)  solution of a conformal
kinetic theory in relaxation time approximation that interpolates between free-streaming $\hat{\gamma}=0$ and perfect fluidity $\hat{\gamma}=\infty$~\cite{Kurkela:2019kip}.
The spatiotemporial temperature profile is given by
\begin{equation}
    T( \tau, \vec x_\perp)= T_{*} \bar  T (\tau/R, \vec x_\perp/R ) \theta(T - T_{\rm F}),\,
    \label{Tevolution}
\end{equation}
where $\bar T$ is a scale invariant solution of the kinetic theory and dimensionful constants $T_{*}$ and $R$ define the temperature normalization and radial size of the system. For different
centrality classes and collision systems the radius $R$ is calculated from 
the entropy density profile $s(x_\perp)$, which we obtain from the
TrENTo initial state model~\cite{Moreland:2014oya}
\begin{align}
    R^2= \frac{\int d^2 x_\perp  (\vec x_\perp-\langle \vec x_\perp \rangle)^2  s(\vec x_\perp)}{\int d^2 x_\perp  s(\vec x_\perp)}\,.
\end{align}
Furthermore, we fixed the temperature normalization $T_*$ to reproduce the centrality dependence of the total entropy $dS/dy=\int d^2 x_\perp s(x_\perp)$, i.e.,
\begin{align}
    T_{*} \propto \left( \frac{dS/dy}{R^2} \right)^{1/3}\, .
    \label{eq:tempvsS}
\end{align}
As a reference value, we choose to set the temperature at the origin in 0-10\% PbPb collisions at time $\tau_\text{ref} = 0.6\,\text{fm}/c$ to be 
$T(\tau_\text{ref}, 0) = 485\,\text{MeV}$ (corresponding to a typical temperature in hydrodynamic simulations of 0-10\% PbPb collisions at $\sqrt{s_{NN}}=5.02\,\text{TeV}$).
We note that none of the predictions of our models depend on the specific choice of $T(\tau_\text{ref},0)$ as it can be reabsorbed in the quenching parameter $\hat{\bar{q}}$. 
The $\theta$-function in  \Eq{Tevolution} implements the model assumption that the medium modifications of hard partons cease at freeze-out at
$T_{\rm F}=175\,\text{MeV}$.
We include interactions between hard partons and the medium
for $\tau > \tau_0 = 0.05\,\text{fm}/c$.
Kinetic solution $\bar T$ is given for times $\tau\gtrsim 0.06R$, so if needed the temperature is back-extrapolated to $\tau_0=0.05\,\text{fm}/c$ using $\tau^{-1/3}$ scaling.
The centrality dependencies of $dS/dy$ and $R$ are tabulated in the \app{sec:geom}. We choose the kinetic theory solution with an opacity $\hat \gamma=16$ which corresponds to an almost perfect ($\eta/s \approx 1/4\pi$) fluid in central $\sqrt{s_{NN}}=5.02\,\text{TeV}$ PbPb collisions~\cite{Kurkela:2019kip}. We compare this fluid limit to the case of free streaming (opacity $\hat\gamma = 0$).

In addition to the azimuthally symmetric profile \Eq{Tevolution}, we model the elliptical deformation of the background profile in off-central nucleus-nucleus collisions. This is achieved by adding a linearized kinetic theory solution of an elliptic background perturbation~\cite{Kurkela:2019kip}. The magnitude of such deformation is fixed by the eccentricity in the initial conditions (see  \app{sec:geom}).

The above formulation of background evolution clearly
aims at simplicity rather than completeness.
However, we checked by drastically changing the temperature evolution in \Eq{Tevolution} that the main conclusions about the system size dependence of the nuclear modification factor \Eq{RAA} do not change significantly (see \Sec{sec:robustness}).
Of course, this does not mean that other observables are not sensitive to these details (see \Sec{sec:v2}), but we leave a more refined description of the background evolution to future works.

\subsection{Embedding hard partons in a medium}
\begin{figure}
    \centering
    \includegraphics[width=\linewidth]{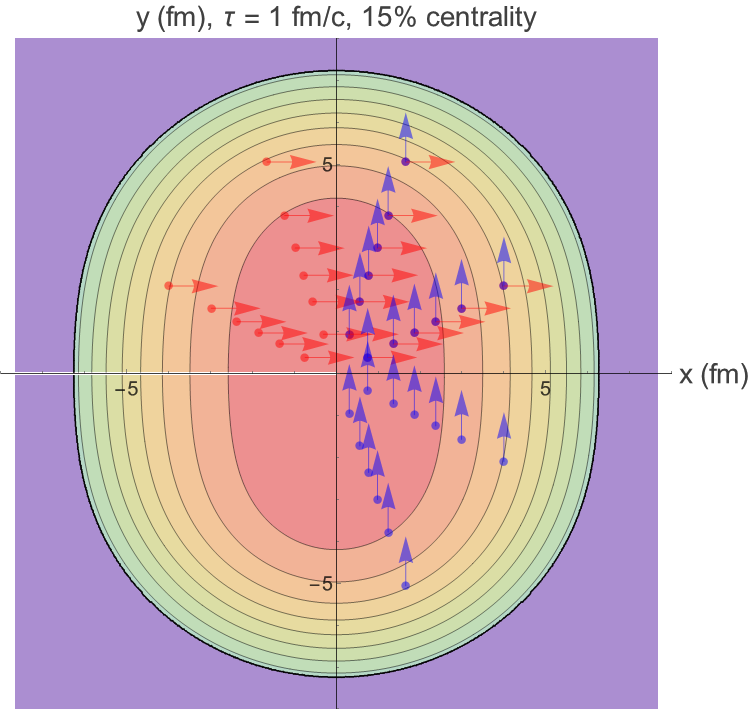}
    \caption{A typical background temperature profile at 15\% centrality at $\tau = 1\,$fm/c. The arrows correspond to the starting location and direction of the sampled partons used to determine the nuclear modification factor, \Eq{RAA}.}
    \label{fig:geometry}
\end{figure}

The quenching parameter $\hat{\bar q}$ is determined by the temperature
profile along  the trajectory $\vec x(t)$ of a particular particle: 
\begin{equation}
	\hat{\bar q}(t ,\vec x(t)) = d \,  \left[ T( t, \vec x(t)) \right]^3\, .
	\label{defqhat}
\end{equation}
Here, the proportionality factor $d$ is a model parameter that will be adjusted to reproduce the medium induced suppression of
single inclusive hadron spectra in central PbPb collisions at $p_\perp \approx 50\,\text{GeV}$ [we keep  $\alpha_s=0.3$ constant in \Eq{dI}].
It is \Eq{defqhat} that relates the modeling of the QCD evolution and the geometrical embedding of parton trajectories in that
medium to the actual dynamics of parton energy loss.

Hard partons are assumed to be produced in binary scatterings and to follow eikonal trajectories in the plane transverse to the beam
\begin{equation}
\vec x(t) = \vec x_0 + \vec v t\, ,\quad  \hbox{with}\quad  v^2 =1\, .
\label{traj}
\end{equation}
For boost invariant medium evolution we can always find such a frame.
The distribution of production vertices $\vec x_0 $  is set to reproduce the (hard) rms radius $R_h$ of binary nucleus-nucleus collisions 
obtained from the product of the nuclear thickness functions of the two nuclei in the TrENTo model (see \app{sec:geom}).
We discretize the velocity angle and initial radial location of the hard particles as shown in \Fig{fig:geometry}. A linear grid in 
radial coordinate $\rho$ with
\begin{align}
    \rho &=1 - \exp(-(r/R_h)^2)
\end{align}
leads to a Gaussian distribution of hard particles in the physical $r$ coordinate.
The values of $R_{h}$ and $\left<N_{\rm coll}\right>$  are documented in the \app{sec:geom}.
For each collision system and centrality, the nuclear modification factor  \Eq{RAA} is obtained by averaging the energy loss of hard partons over the ensemble of starting locations and velocities shown in \Fig{fig:geometry}. 
We obtain minimum bias results by taking the $N_{\rm coll}$-weighted average over 10 centrality classes.

\subsection{Vacuum parton and hadron spectra}
\label{sec:vacuum}

In the absence of parton energy loss, the single inclusive hadron (parton)  spectra can be calculated in collinearly factorized perturbative QCD according to \Eq{schematicv2}. For the proton reference spectrum, we take PDFs provided by CT14~\cite{Dulat:2015mca} and for oxygen and lead nuclei we use nPDFs derived from EPPS16 global fit~\cite{Eskola:2016oht}. We convolute the PDFs with LO QCD scattering matrix elements to produce the vacuum spectra $d\sigma^\text{vac}_{g/q}$ of quarks and gluons (for the nuclear modification factor, the difference between LO and next to leading order results is negligible~\cite{Huss:2020dwe}).
The charged hadron cross section is  obtained from the partonic one by the convolution with the quark and gluon fragmentation functions $D^{g/q}_h$ using Binnewies-Kniehl-Kramer (BKK) parametrization~\cite{Binnewies:1994ju}
\begin{align}
     \frac{d \sigma^{h,\text{vac}}_{g/q}}{d p_\perp^2} = \int_0^1 \frac{dz}{z^2} D^{g/q}_h(z) \frac{d \sigma^\text{vac}_{g/q}(p_\perp/z)}{d p_\perp^2}\, ,
     \label{hadspec}
\end{align}
where $z$ is the momentum fraction of the parton that is carried by the leading hadron. We use the LHAPDF6 interpolator for evaluating PDFs and FFs~\cite{Buckley:2014ana}.
Details of the computation are summarized in the \app{sec:cross}. 

\begin{figure}
    \centering
    \includegraphics[width=\linewidth]{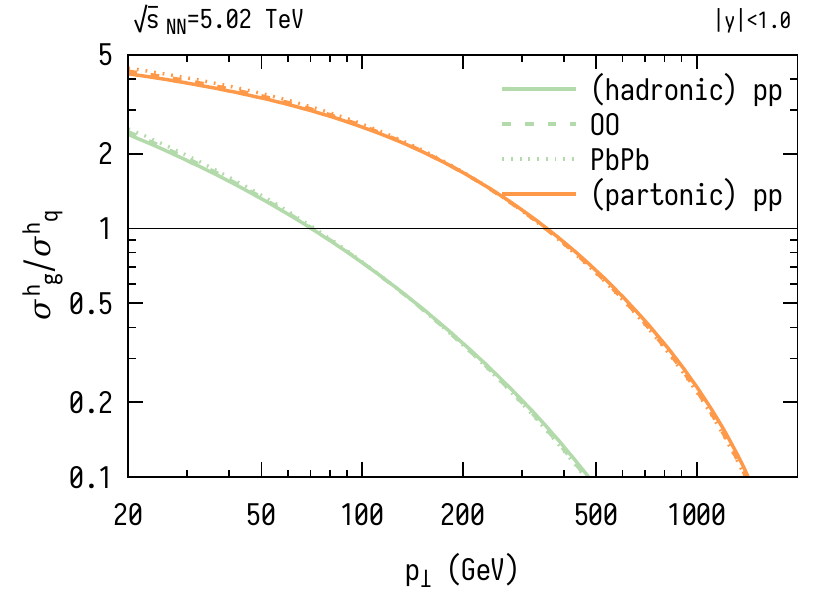}
    \caption{Green lines show the ratio of gluon and quark contributions to  the inclusive charged hadron cross section in $pp$, OO and PbPb collisions at $\sqrt{s_{NN}}=5.02\,\text{TeV}$. Orange lines show the corresponding gluon to quark ratios before fragmentation. }
    \label{fig:hadfrac}
\end{figure}

In \Fig{fig:hadfrac} we show the ratio of quark and gluon fragmentation contributions to the inclusive charged hadron (parton) cross section at $\sqrt{s_{NN}}=5.02\,\text{TeV}$ for different collision systems, i.e.
\begin{equation}
    r(p_\perp) = \frac{d\sigma^{h,\text{vac}}_{g}/d^2p_\perp}{d\sigma^{h,\text{vac}}_q/dp^2_\perp}\, .\label{eq:r}
\end{equation}

\begin{figure}
    \centering
    \includegraphics[width=\linewidth]{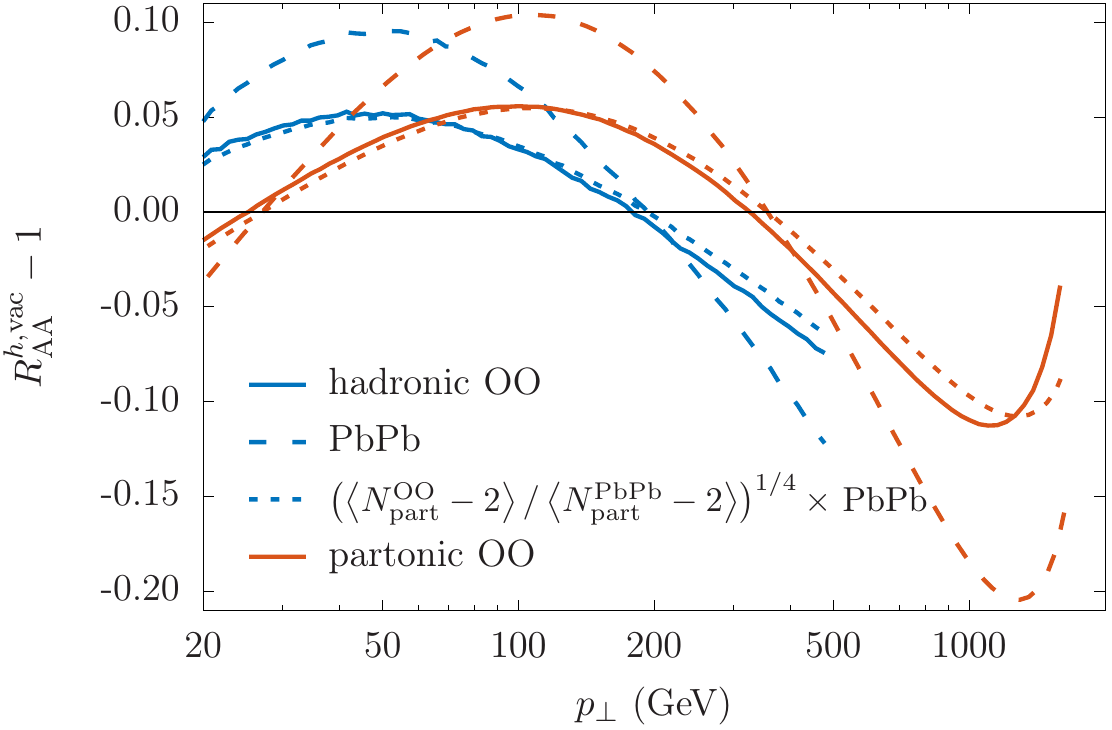}
    \caption{The blue lines show the hadron nuclear modification factor \Eq{RAAvac} for OO ($A=16$) and PbPb ($A=208$) collisions in the absence of parton energy loss. Deviations from unity indicate nPDF effects (nPDF uncertainties not shown). We also show rescaled PbPb modification with number of participant nucleons, where $\left< N_\text{part}^\text{OO}\right>\approx 10.4$ and $\left< N_\text{part}^\text{PbPb}\right>\approx 114$. Red lines show the corresponding partonic nuclear modification factors before fragmentation.
    }
    \label{fig:RAAnpdf}
\end{figure}

Although gluons dominate the partonic spectra at momenta up to  $p_\perp\approx 300\,\text{GeV}$, they fragment to softer hadrons than quarks and therefore the hadron spectrum is dominated by quark fragmentation already at $p_\perp>70\,\text{GeV}$.  $r(p_\perp)$ does not change significantly between  $pp$ and AA collisions (in the absence of energy loss), although the  nPDF modifies the absolute yields. We computed such ``vacuum" nuclear modification factor
\begin{equation}
     R_\text{AA}^{h, \text{vac}}(p_\perp) =\frac{1}{A^2} \frac{d\sigma^{h,\text{vac}}_\text{AA}/dp^2_\perp}{d\sigma^{h,\text{vac}}_{pp}/dp^2_\perp}
     \label{RAAvac}
\end{equation}
for hadrons and partons in OO and PbPb collisions, see \Fig{fig:RAAnpdf}. We emphasize that here we take the central values of nPDFs~\cite{Eskola:2016oht}. Within current  nPDFs uncertainties, the modifications shown in  \Fig{fig:RAAnpdf} are consistent with zero for most of the kinematic range. Taking into account such uncertainties (and constraining them with further data) is crucial for disentangling the different sources of nuclear modification in comparison to experimental data. We address this question in detail in our companion paper~\cite{Huss:2020dwe}, so we will not discuss nPDF uncertainties further here.

We see that nPDF effects become smaller with decreasing $A$.  We find empirically that the nPDF contribution to the nuclear modification scales well with 
 $(\left<N_\text{part}\right>-2)^{1/4}$, where $\left<N_\text{part}\right>$ is the average number of participant nucleons. 
As nPDF effects are expected to be smaller in peripheral collisions~\cite{Helenius:2012wd}, we use our empirical  scaling to estimate the nPDF effects in centrality selected events. For each centrality class we take this factor to be 
\begin{equation}
\frac{\left. R_\text{AA}^{h, \text{vac}}(p_\perp)\right|_\text{cent} - 1}{ \left. R_\text{PbPb}^{h, \text{vac}}(p_\perp)\right|_\text{min bias} - 1}= k \left(\left.\left<N_\text{part}\right>\right|_\text{cent}-2\right)^{1/4},\label{eq:RAAcent}
\end{equation}
where $k=0.25$ is a normalization such that for PbPb the $N_\text{coll}$-weighted centrality average reproduces the minimum bias nuclear modification factor.

\subsection{System size dependence of parton energy loss}
\label{kinksec}

\begin{figure}[t]
    \centering
    \includegraphics[width=\linewidth]{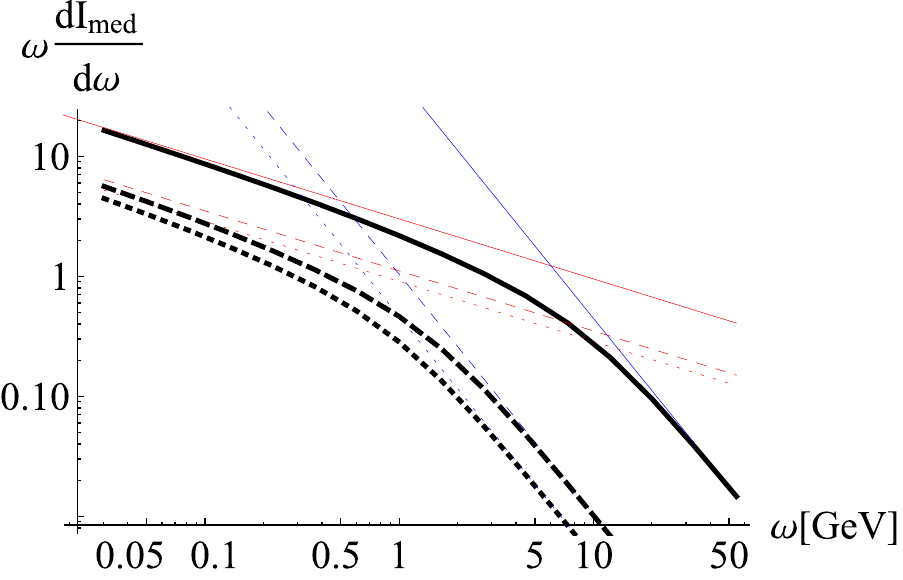}
    \includegraphics[width=\linewidth]{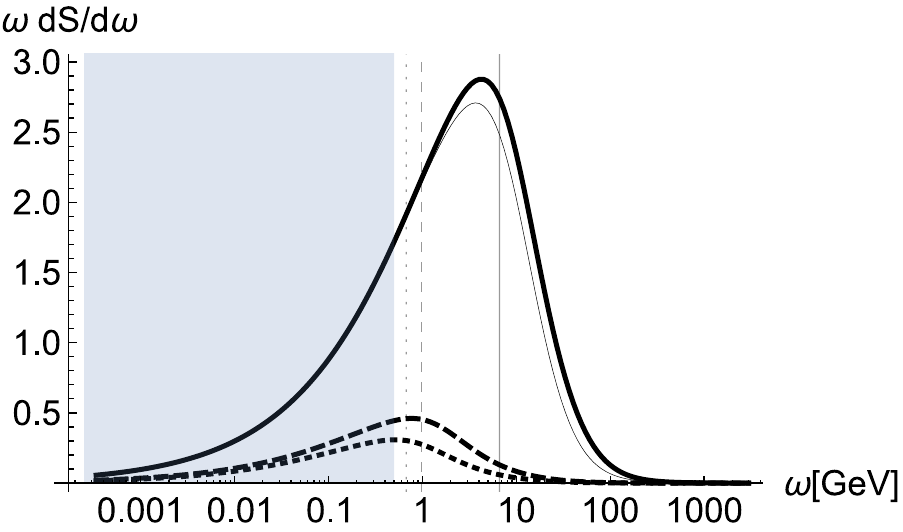}
    \caption{(top) The probability of medium-induced bremsstrahlung $\omega dI^g_\text{med}/d\omega$ for a hard gluon starting from the center of the collision system for PbPb (solid lines), OO (dashed lines), and $p$Pb (dotted lines). The blue (steeper) and the red (more gradual) lines correspond to asymptotic solutions \Eq{opacity} and \Eq{LPM} respectively. The frequency where the two asymptotic rates are equal defines $\omega_{\rm kink}$ that is for these systems approximately at $\omega^{\rm PbPb}_\text{kink} \approx 6.6\,\text{GeV}$, $\omega^{\rm OO}_\text{kink} \approx 1.0\,\text{GeV}$, and  $\omega^{p\text{Pb}}_\text{kink} \approx 0.7\,\text{GeV}$.
    (Here we have chosen $\hat{\bar{q}}/T^3= 2.46$.).
    (botom panel) Integrand of the shift function  \Eq{Sint}. The area under the curves represents contributions to $S_g$ by gluon emission at different energy  scales for the final hadron with $p_\perp = 100\,\text{GeV}$ (thin line, $p_\perp = 50\,\text{GeV}$) and $\left<z n\right> \approx 3$. The vertical lines correspond to $\omega_{\rm kink}$; the shaded region corresponds to $\omega < 500\,\text{MeV}$.}
    \label{fig:rateplot}
\end{figure}

For any generic quenching parameter \Eq{defqhat} associated to a particular parton trajectory \Eq{traj} through a QCD medium of 
given temperature profile \Eq{Tevolution}, we can solve numerically the differential equation \Eq{Arnolddiff}
and we can thus determine the medium-modified gluon energy distribution $\omega \textstyle\frac{dI^g_{\rm med}}{d\omega}$
in  \Eq{dI}. For trajectories starting in the center of central
PbPb, OO and $p$Pb collisions, the resulting medium induced gluon rates $\omega \textstyle\frac{dI^g_{\rm med}}{d\omega}$ are illustrated
in the top panel  of Fig.~\ref{fig:rateplot}. The main qualitative characteristics of these numerical results can be understood 
by considering the following limiting cases~\cite{Arnold:2008iy}:
\begin{enumerate}
\item For transparent systems, i.e., small $\hat{\bar{q}}$, \Eq{Arnolddiff} can be solved iteratively around the vacuum solution $c_\text{vac}(t)=1$,
\begin{align}
     \ln|c(0)| = \frac{1}{2}|c_1(0)|^2 + c_2(0),
\end{align}
where
\begin{align}
    c_1(t) &= i \frac{C_A}{2\omega} \int_t^{\infty}  dt'\,  (t'-t)  \hat{\bar{q}}(t' , \vec x(t' )) \, , \\
     c_2(0) &=  i \frac{C_A}{2\omega} \int_0^{\infty}  dt\,  t  \hat{\bar{q}}(t , \vec x(t ))  c_1(t )\, .
\end{align}
The resulting emission rate is
\begin{align}
    \omega\frac{d I^\text{transp.}_\text{med}}{d \omega}  &\propto \frac{\alpha_s}{\omega^2}\left[ \int_0^\infty dt\, t \hat{\bar{q}}(t,\vec x(t))  \right]^2.
    \label{opacity}
\end{align}
\item 
For large (opaque) slowly varying systems with $|\dot \omega_0(t)|\ll |\omega_0^2(t)|$, \Eq{Arnolddiff} can be solved using adiabatic approximation $c(t)\approx \exp\left[ i\int_t^\infty dt' \omega_0(t')\right]$. The solution is
\begin{align}
    \ln|c(0)| = \frac{\sqrt{C_A}}{2\sqrt{\omega}}\int dt \sqrt{\hat{\bar{q}}(t,\vec x(t))}
    \label{eq:themralRate}
\end{align}
for which
\begin{align}
   \omega \frac{d I^\text{opaq.}_\text{med}}{d \omega}  &\propto \frac{\alpha_s}{\sqrt{\omega}}\int_0^\infty dt \sqrt{\hat{\bar{q}}(t,\vec x(t)) } \, .
    \label{LPM}
\end{align}
\end{enumerate}
Comparing the parametric estimates  \Eq{opacity} and \Eq{eq:themralRate}, one finds that
the crossover between these two limiting cases occurs at a frequency $\omega_{\rm kink}$ 
\begin{align}
    \omega_{\rm kink} \propto \frac{\left[ \int_0^\infty dt\, t\, \hat{\bar{q}}(t, \vec x(t))  \right]^{4/3}}{\left[ \int_0^\infty dt \sqrt{\hat{\bar{q}}(t, \vec x(t)) }\right]^{2/3}}\, .
    \label{wkink}
\end{align}

In the upper panel of Fig.~\ref{fig:rateplot} we illustrate the characteristic interpolation between the non-Abelian Landau-Pomeranchuk-Migdal
(LPM) $\omega^{-1/2}$-powerlaw of \Eq{eq:themralRate} in the limit of soft gluon energies, $\omega \ll  \omega_{\rm kink} $
and the $\omega^{-2}$ powerlaw \Eq{opacity} of the opacity expansion for $\omega \gg  \omega_{\rm kink} $. As the integrals in 
\Eq{wkink} depend on the in-medium path length and the density of the system, $\omega_{\rm kink}$ depends on the 
QCD medium produced in the collision and is larger for systems of larger geometrical extent and/or larger density (see caption of 
\Fig{fig:rateplot} for numerical details).

\subsection{Quenching of the hadron spectrum}
\begin{figure}
    \centering
    \includegraphics[width=\linewidth]{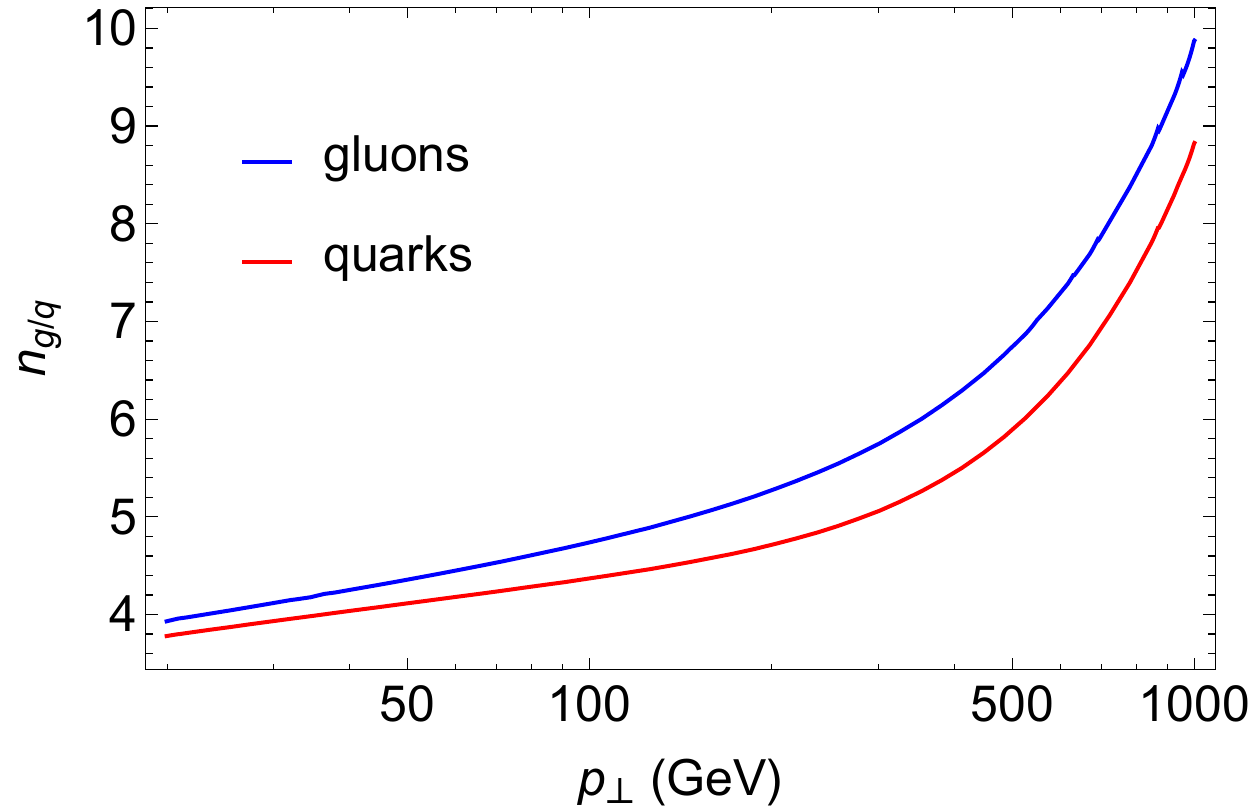}
    \includegraphics[width=\linewidth]{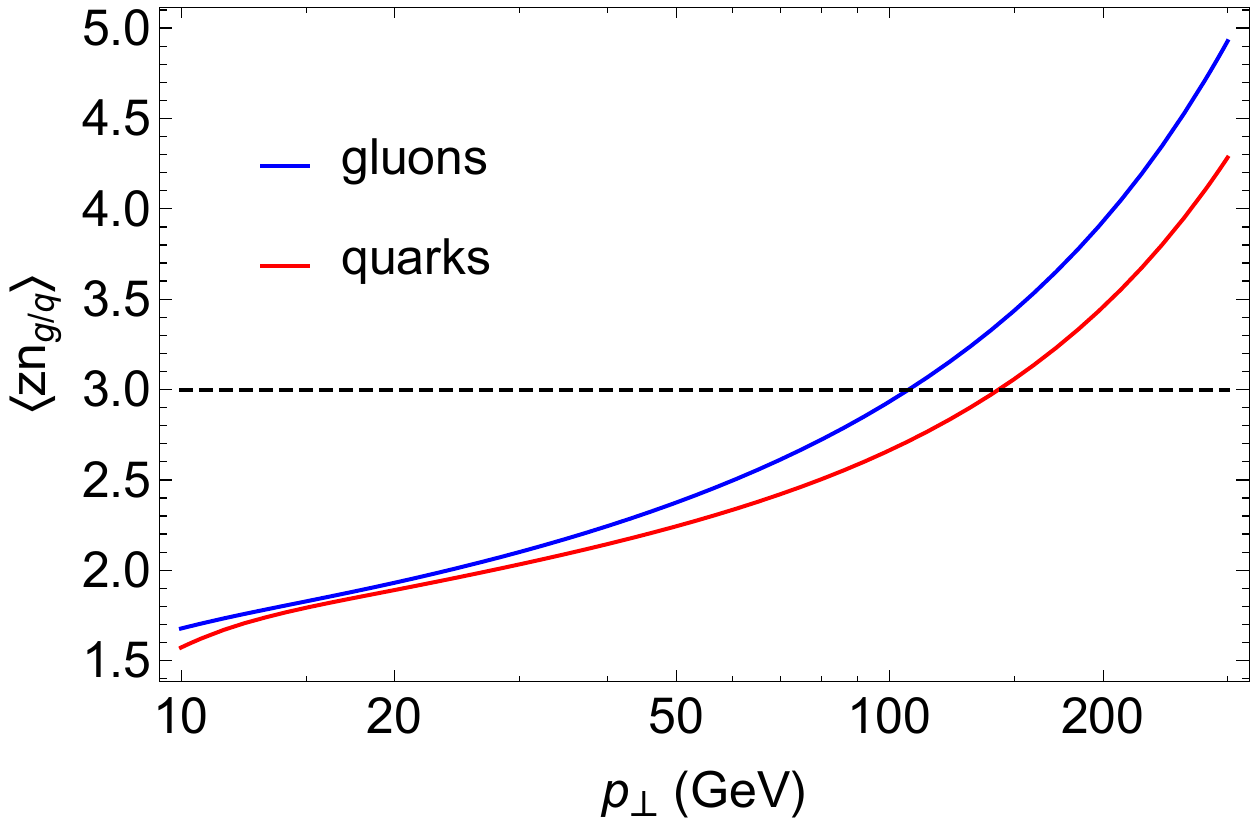}
    \caption{(top) Spectral index for gluons (upper curves) and quarks (lower curves), \Eq{eq:n}. (bottom) The reduced exponent \Eq{zn} for fragmented gluons and quarks.}
    \label{fig:neff}
\end{figure}

Having calculated for each trajectory \Eq{traj} the medium-induced gluon rate $\omega \frac{dI_{\rm med}}{d\omega}$ as 
illustrated in the upper panel of Fig.~\ref{fig:rateplot}, we determine the corresponding 
probability $P(\epsilon)$ of parton energy loss in \Eq{quenchweight}. To characterize the impact of parton energy loss, 
we consider  the ratio of partonic medium modified and vacuum cross sections, i.e., the quenching factor~\cite{Baier:2001yt}
\begin{eqnarray}
    Q_{g/q}(p_\perp) &=& \frac{d \sigma^\text{med}_{g/q}(p_\perp)/dp_\perp^2}{d \sigma^\text{vac}_{g/q}(p_\perp)/dp_\perp^2} 
    \nonumber\\
     &=& \int d\epsilon P_{g/q}(\epsilon ) \frac{d \sigma^\text{vac}_{g/q}(p_\perp+\epsilon)/dp_\perp^2}{d \sigma^\text{vac}_{g/q}(p_\perp)/dp_\perp^2}  \, .
     \label{Qgq}
\end{eqnarray}
For $\epsilon \ll p_\perp$ we can approximate\footnote{
We employ an alternative rewrite of the Taylor series 
\begin{equation}
    f(x) =\exp \sum_{n=0}^\infty \frac{x^n}{n!}\frac{\partial^n \log f(x)}{\partial^n x}.
\end{equation}
}
\begin{align}
    Q_{g/q}(p_\perp)&\approx \int d\epsilon 
    P_{g/q}(\epsilon) e^{- n_{g/q}(p_\perp) \epsilon/p_\perp}\, .
    \label{Qgp}
\end{align}
where $n_{g/q}(p_\perp)$ is the spectral index 
\begin{equation}
    n_{g/q}(p_\perp) = -\frac{d \log (d \sigma^\text{vac}_{g/q}(p_\perp)/dp_\perp^2)}{d\log p_\perp}.\label{eq:n}
\end{equation}
Note that partonic spectra are falling steeply with   $n\gtrsim 5$  in the kinetic regime $20\, {\rm GeV} < p_\perp < 1000\, {\rm GeV}$ relevant for our study, see \Fig{fig:neff}.

In close analogy to \Eq{Qgq}, we define also the suppression of charged hadrons due to parton energy loss 
by the ratio
\begin{align}
    Q^h_{g/q}(p_\perp) = \frac{d \sigma^\text{med}_{h,g/q}(p_\perp)/dp_\perp^2}{d \sigma^\text{vac}_{h,g/q}(p_\perp)/dp_\perp^2}\, ,
    \label{Qh}
\end{align}
where $\sigma^\text{vac}_h$ is the single inclusive charged hadron cross section in vacuum,  and $ \sigma^\text{med}_h$ is the corresponding quantity with medium  included modifications.
Fragmented hadrons are produced at softer momenta, which results in the partonic cross section effectively changing momentum by a factor of $z\approx 0.5$, i.e., $ \sigma^\text{vac}_{g/q}(p_\perp)/dp_\perp^2\approx \sigma^\text{vac}_{h,g/q}(0.5p_\perp)/dp_\perp^2$.
Analogously  we can write the result in exponential form \Eq{Qgp} with reduced exponent\footnote{The approximation amounts to assuming $\langle z^n\rangle = \langle z\rangle^n$. By doing the fragmentation of the quenched partonic spectra in \Eq{Qh} directly, we verified that this does not qualitatively alter the nuclear modification factor.}
\begin{equation}
   \left<zn_{g/q}\right>(p_\perp) = \frac{\int_0^1 \frac{dz}{z^2} D^{g/q}_h(z) z n_{g/q}(p_\perp/z) d \sigma^\text{vac}_{g/q}(p_\perp/z)/dp_\perp^2}{\int_0^1 \frac{dz}{z^2} D^{g/q}_h(z) d \sigma^\text{vac}_{g/q}(p_\perp/z)/dp_\perp^2}.\label{zn}
\end{equation}
where typically  $\left<zn\right>\approx 3$. 
 In \Fig{fig:neff} we display the momentum dependence of $\left<zn_{g/q}\right>$ for hadrons produced by quark and gluon fragmentation.

The exponential form of \Eq{Qgp} allows for a particularly simple evaluation of the integral over the probability distribution \Eq{quenchweight}.
For large hadron momentum the medium modification of the hadron spectra is proportional to the mean energy loss
\begin{align}
    Q^h_{g/q}(p_\perp)&\approx 1 -  \frac{\left<z n_{q/g}\right>}{p_\perp}\left<\epsilon\right>+\ldots\label{Qgp2}
\end{align}
For generic $p_\perp$, the result can be expressed with a shift function $S_{g/q}(u)$  as
\begin{equation}
 	Q^h_{g/q}(p_\perp) = \exp\left[-\frac{\left<z n_{q/g}\right>}{p_\perp} S_{g/q}\left({\left<zn_{g/q}\right>}/{p_\perp}\right) \right]\, ,
 	\label{eq:Qzn}
\end{equation}
where $S_{g/q}(u)$ 
denotes the energy loss
due to multiple medium-induced
gluon emissions~\cite{Baier:2001yt}
\begin{align}
S_{g/q}(u)&= \frac{1}{u}\log \int_0^\infty d\epsilon P_{g/q}(\epsilon) e^{-u\epsilon}\nonumber\\
&=\int_0^\infty d\omega  \frac{ 1- e^{- u \omega}}{u\omega} \omega\frac{d I^{g/q}_{\rm med}(\omega)}{d\omega}.
\label{Sint}
\end{align}
As discussed in \Sec{kinksec}, the characteristic emission  energy  $\omega\frac{dI^{g/q}_{\rm med}}{d \omega}$  has a UV cutoff at $\omega_{\rm kink}$, \Eq{wkink},
therefore if $u\omega_\text{kink}\ll 1$ (which is usually the case), 
the  energy loss \Eq{Sint} becomes proportional to the integral over the gluon 
emission rate $\omega \frac{dI_{\rm med}}{d\omega}$. 
From \Eq{LPM} one finds for the quenching weight the parametric form 
\begin{align}
\log Q^h_{q/g} \propto -\alpha_s \frac{\left<zn_{g/q}\right>}{p_\perp}\sqrt{\omega_{\rm kink}} \int_0^\infty dt \sqrt{\hat{\bar{q}}(t,\vec x(t))}\, .\label{eq:logQ}
\end{align}
In the following, the quenching factor  will be calculated  using the full integral \Eq{Sint}.

Finally, the hadron nuclear modification factor can be computed by multiplying the nPDF modification, \Eq{RAAvac}, with appropriately weighted quark and gluon quenching factors for the hadron spectra 
\begin{equation}
  R_\text{AA}^{h} = \frac{R_\text{AA}^{h,\text{vac}}(p_\perp)}{1+r(p_\perp)} \left[Q_q^h(p_\perp) +   Q_g^h(p_\perp) r(p_\perp)\right],\label{eq:RAAfinal}
\end{equation}
where the $r(p_\perp)$ ratio is given by \Eq{eq:r}.

\subsection{Model applicability in small collision systems}

Parton energy loss models have been applied so far to relatively large collision systems. Here we ask whether the parametric
range of applicability of the parton energy loss model extends to smaller systems like inclusive OO collisions or even $p$Pb
collisions.

The BDMPS-Z formalism was developed for the emission of sufficiently energetic gluons to which a perturbative reasoning applies.
To establish to what extent this condition is met in our model calculations, we show in the lower panel of Fig.~\ref{fig:rateplot}
the integrand of the energy loss function \Eq{Sint} for typical model parameters of collision systems of different size, and for 
typical hadronic transverse momenta $p_\perp = 50$ and $100$ GeV considered in the following. 
The integrand of \Eq{Sint} depends only weakly on $p_\perp$ in the kinematical range of phenomenological
interest, and the scale $\omega_{\rm kink}$ is seen to characterize the peak of the integrand for all collision systems.
 In calculations we consistently assumed $\omega \ll p_\perp$, which is approximately fulfilled for $p_\perp > 50\,\text{GeV}$ in the largest collision systems and holds for much lower momentum in smaller systems.
The characteristic energy of  medium induced gluon radiation
$\omega_{\rm kink}$ decreases with decreasing density and geometric extent of the system,
and the integral \Eq{Sint} receives an increasing contribution 
from very soft gluon emission for which the validity of our model becomes questionable.
We note however that the extrapolation to small systems shown in \Fig{fig:rateplot} is smooth and roughly half of the computed energy loss can be attributed to radiation with $\omega\gtrsim 1\,\text{GeV}$ for OO collisions. 
With these considerations we take a pragmatic approach of
basing a first exploratory study of the systems size dependence of  parton energy loss on a BDMPS-Z formalism that is
not modified with additional assumptions for small systems.

We mention as an aside that we have performed other consistency checks of our model setup. In particular, the
discussion above assumed $x \ll 1$. We checked that relaxing this approximation has only mild effects on the results
in Fig.~\ref{fig:rateplot} (data not shown). Within the model uncertainties quoted in the present paper, these are negligible,
and we do not discuss them further. We also checked that the phenomenological practice of mapping parton energy loss of a
dynamically evolving QCD medium onto a parton energy loss calculation for a static brick of suitably chosen quenching
parameter describes, over the entire $\omega$-range, the energy loss curve in Fig.~\ref{fig:rateplot} within 5\% accuracy. We
do not employ this observation to simplify our calculation, but we note it here since it implies that our results could be reproduced
in other existing approaches. 

\begin{figure}
    \centering
     \includegraphics[width=\linewidth]{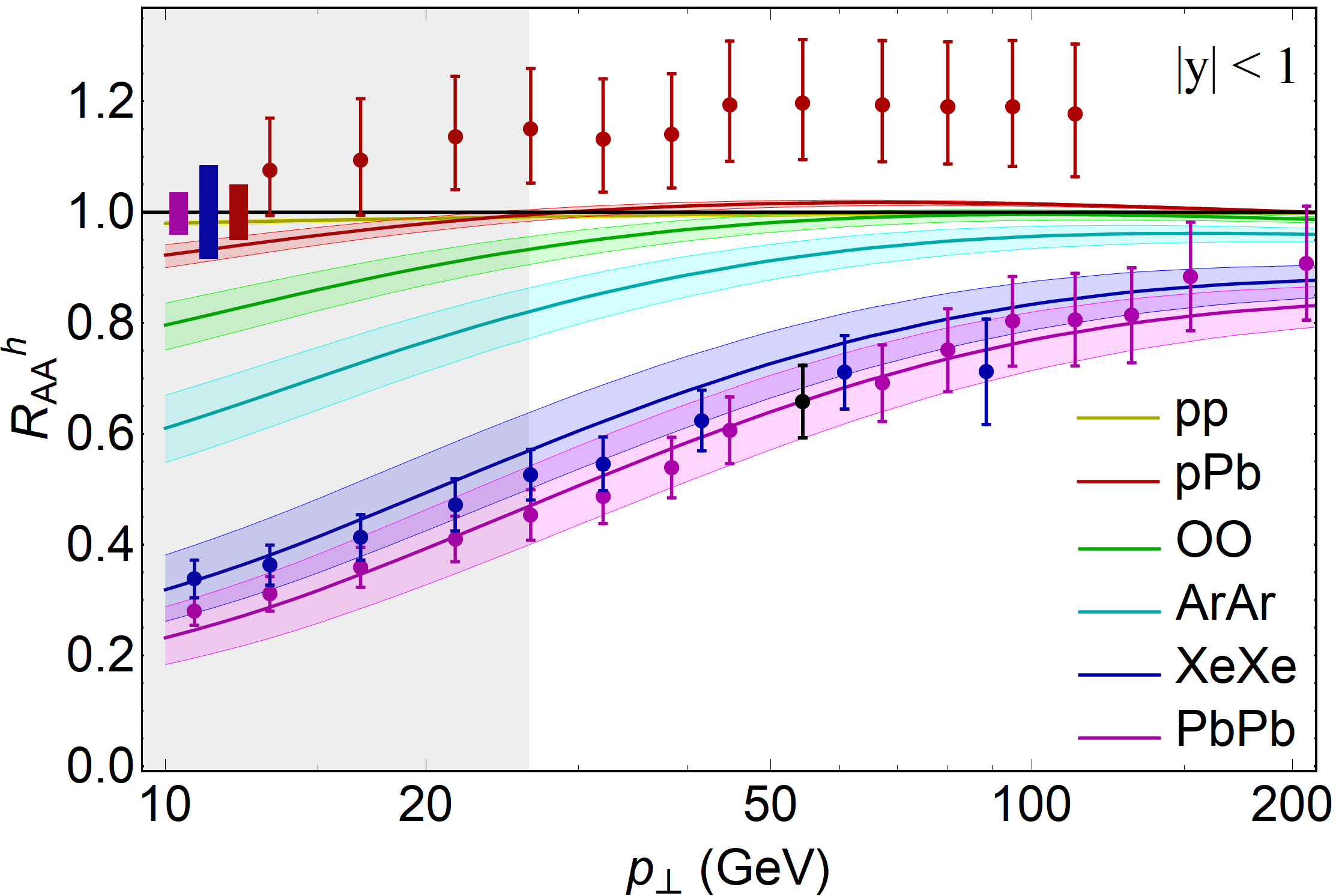}
    \caption{The nuclear modification factor $R^h_\text{AA}$ for different centrality averaged collision systems (curves follow the ordering of the legend).  Normalization uncertainties in  PbPb, XeXe and $p$Pb data are shown as boxes~\cite{Khachatryan:2016odn,Sirunyan:2018eqi}.
    }
    \label{fig:MinBiasPlot}
\end{figure}

\begin{figure}
    \centering
     \includegraphics[width=\linewidth]{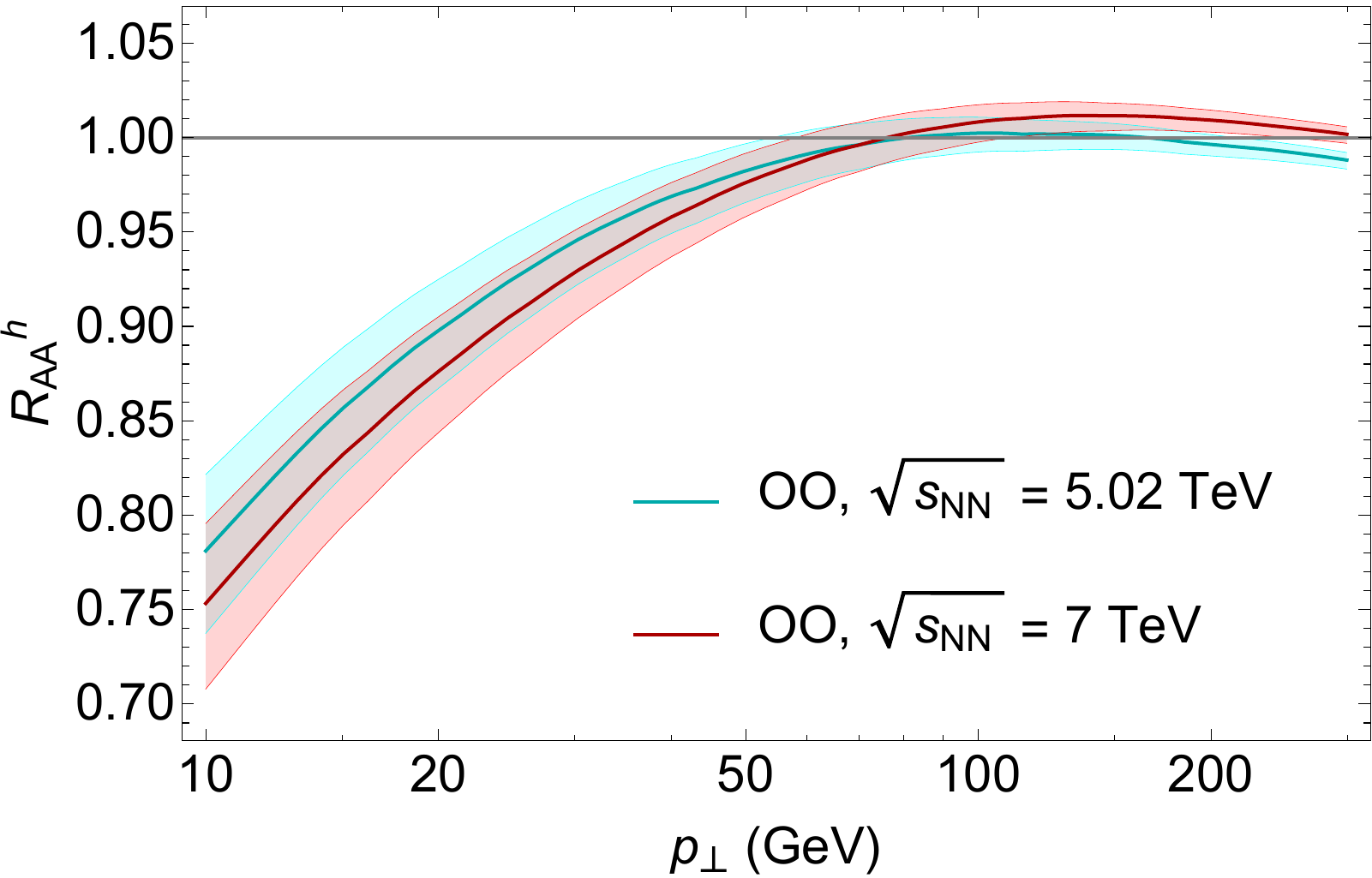}
    \caption{
    Comparison of the minimum bias hadron nuclear modification factor in OO collisions at $\sqrt{s_\text{NN}}=5.02\,\text{TeV}$ (upper band)
and $\sqrt{s_\text{NN}}=7\,\text{TeV}$ (lower band).
    }
    \label{fig:OO7TeV}
\end{figure}

\begin{figure*}
    \centering
        \includegraphics[width=\linewidth]{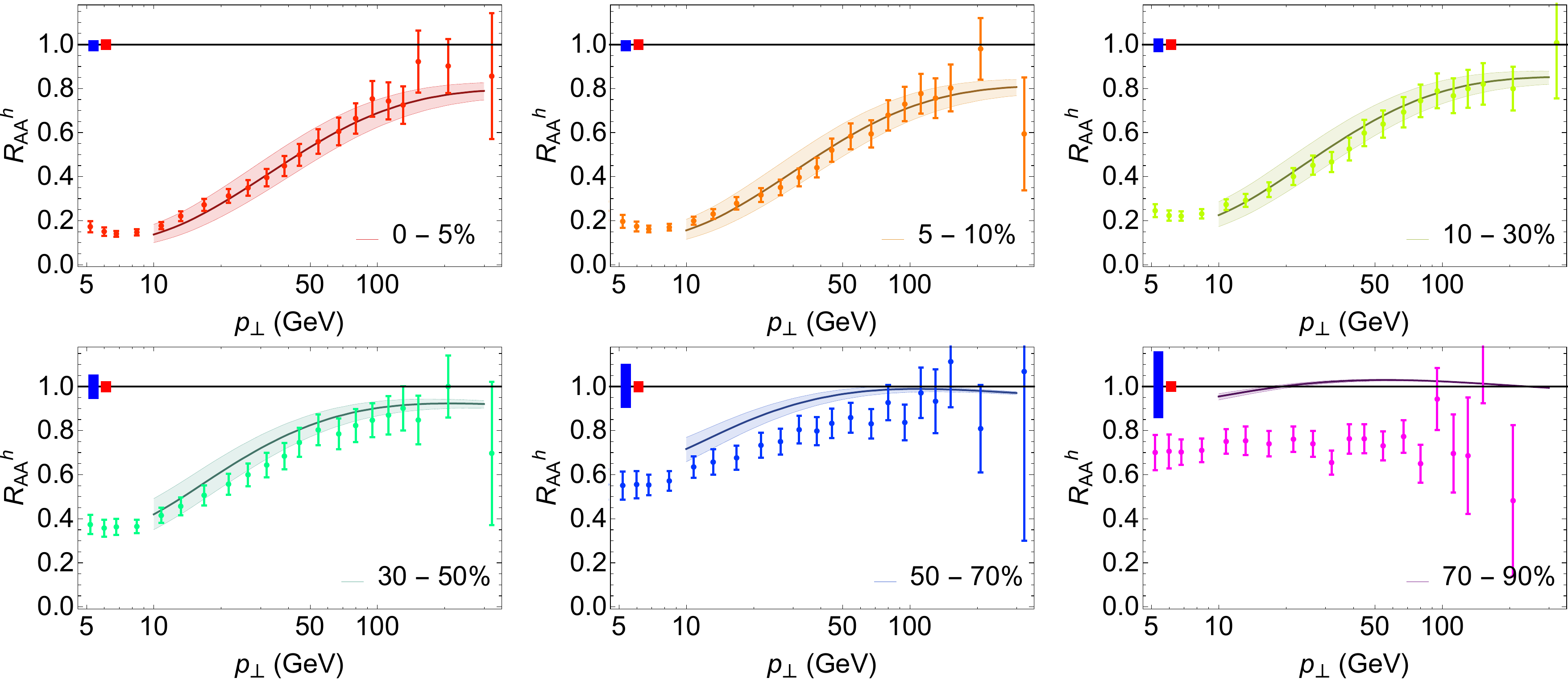}
    \caption{The charged hadron nuclear modification factor
    $R_\text{AA}^h$ in $\sqrt{s} =5.02\,\text{TeV}$ PbPb collisions shown in six centrality bins. Results of the parton energy loss model of Section~\ref{sec2} are compared to data from Ref.~\cite{Khachatryan:2016odn}. Blue (red) boxes indicate systematic experimental uncertainties in nuclear thickness $\langle T_\text{AA}\rangle$ (luminosity) that affect the normalization of $R_\text{AA}^h$.}
    \label{fig:centralityPbPb}
\end{figure*}

\section{Results}
\label{sec3}
We now compare results of the parton energy loss model described above to the measured centrality and momentum dependence of the charged hadron nuclear modification factor $R_\text{AA}^{h}(p_\perp)$ in PbPb and XeXe collisions at the LHC. We then predict the nuclear modification factors in minimum bias $p$Pb, OO and ArAr collisions, and 
centrality selected OO collisions. We test the robustness of these results by varying model assumptions. Finally, we discuss to what extent parton energy loss can account for the observed azimuthal momentum anisotropy $v_2(p_\perp)$ at sufficiently high transverse momentum within our setup.

\subsection{System size and momentum dependence}
\label{sec3a}
If the temperature profile of the QCD medium is fixed,
the only remaining unconstrained parameter of 
the parton energy loss model of Sec~\ref{sec2} 
is the proportionality factor $d$ that sets 
the value of the quenching parameter $\hat{\bar{q}}$ in units of $T^3$ in \Eq{defqhat}. 
We adjust $d$ such that the model reproduces the measured centrality averaged hadron nuclear modification factor  $R_\text{AA}^h(p_\perp = 54.4\, {\rm GeV}) = 0.658\pm0.065 $ in $\sqrt{s_{NN}} = 5.02\,\text{TeV}$  PbPb collisions at the LHC, see Fig.~\ref{fig:MinBiasPlot}. The resulting
central value is $d=\hat{\bar q}/T^3 = 3.63$. Variation of the model parameter in the range $d=\hat{\bar q}/T^3 \in \left[2.72, 4.54\right]$ spans the $R_\text{AA}^h(p_\perp= 54.4\, {\rm GeV})$ values within the $1
$-$\sigma$ experimental uncertainties.

Once the overall normalization of $\hat{\bar{q}}$ is thus fixed, the $p_\perp$-dependence of $R_\text{AA}^{h}(p_\perp)$, its dependence on centrality, and its dependence on the nucleon number $A$ in  centrality averaged collisions are model predictions. 
Fig.~\ref{fig:MinBiasPlot} shows that the model describes well the observed $p_\perp$-dependence in centrality averaged PbPb and XeXe collisions.
Here the error bands account only for the above mentioned variation  of $\hat{\bar{q}}/T^3$.
The same figure also shows
model predictions for minimum bias OO and ArAr collisions at $\sqrt{s_{NN}}=5.02\,\text{TeV}$.

In \Fig{fig:MinBiasPlot} we also compare the same model to measurements of
the nuclear modification factor in $p$Pb collisions. At $p_\perp\approx \mathcal{O}(100)\,\text{GeV}$ the model predicts a slight enhancement of $R^h_{p\text{Pb}}$ indicating that the nuclear modification of the PDFs in the anti-shadowing region is numerically more important than the small parton energy loss~\cite{Eskola:2016oht}. We note that within current theoretical and experimental uncertainties no firm statement about the discrepancy  between data and model predictions for $p$Pb shown in    \Fig{fig:MinBiasPlot} can be made.

Up to now we followed the standard assumption that parton energy loss is negligible in $pp$ collisions. To check the internal consistency of our model we estimated the expected energy loss in $pp$ collisions. The yellow band in
\Fig{fig:MinBiasPlot} shows the ratio of hadron spectra with and without the medium effects. In light of other model uncertainties, this assumption seems justified.
 
 In \Fig{fig:OO7TeV} we show how the nuclear modification factor in centrality averaged OO collisions
 evolves from $\sqrt{s_{NN}}=5.02$ to $7\, \text{TeV}$---the 
 projected center-of-mass energy of the upcoming OO run at the LHC~\cite{Citron:2018lsq}. The effect of changing collision energy is two-fold. First, an increase in $\sqrt{s_{NN}}$ shifts the nPDF effects to higher transverse momentum. Second, the soft medium produced in the collision also depends on the collision energy. Here, we model this by assuming  $T_*\propto s_{NN}^{0.05}$ in \Eq{Tevolution}, which is motivated by the  charged particle multiplicity dependence on center-of-mass energy~\cite{Acharya:2018hhy}.

\begin{figure*}
    \centering
    \includegraphics[width=\linewidth]{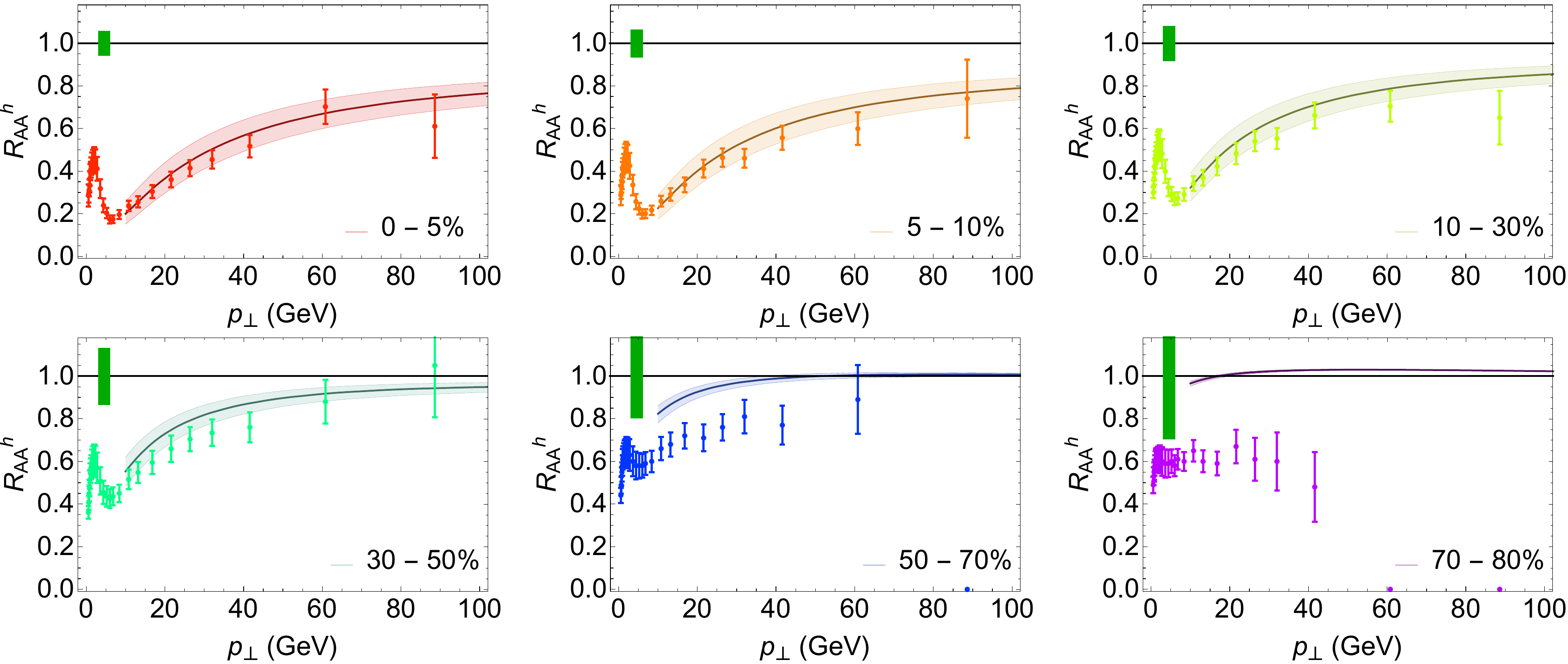}
    \caption{
    The charged hadron nuclear modification factor
    $R_\text{AA}^h$ in $\sqrt{s} =5.44\,\text{TeV}$ XeXe collisions shown in six centrality bins. Results of the parton energy loss model of Section~\ref{sec2} are compared to data from Ref.~\cite{Sirunyan:2018eqi}. 
    The green boxes indicate systematic normalization uncertainty in the measurement of $R_\text{AA}^h$ (as a fraction of  $R_\text{AA}^h$).
    }
    \label{fig:centralityXeXe}
\end{figure*}

In \Fig{fig:centralityPbPb} (\Fig{fig:centralityXeXe}) we compare the $p_\perp$ and centrality dependence of the charged hadron nuclear modification factor in our model and measured data at $\sqrt{s_{NN}} = 5.02\,\text{TeV}$  PbPb  ($\sqrt{s_{NN}} = 5.44\,\text{TeV}$ XeXe) collisions.
The $p_\perp$ dependence of $R_\text{AA}^h(p_\perp)$
mainly stems from the steeply falling particle spectra, while
the centrality dependence is driven by the in-medium path length, see \Eq{eq:logQ}.
As seen in \Fig{fig:centralityPbPb} and \Fig{fig:centralityXeXe}, 
the model reproduces without any parameter adjustment both
the $p_\perp$ and 
centrality dependence of $R_\text{AA}^h$ between 0 and 70\%. At very high $p_\perp$ the fractional energy lost by the parton is small and $R_{\text{AA}}^h$ is dominated by nPDF effects.
We note that systematic normalization uncertainties in the experimental data are shown by blue (green) boxes, which increase to $\approx 15\%$ ($\approx 30\%$) in the most peripheral bin. If these are taken into account, the tension between data and model results visible in the 70-90\% (70-80\%) centrality bin lies within the 2-$\sigma$ uncertainty band.  
We note however that no parton energy loss model of BDMPS-Z type contains physics that could account for a stagnation or an increase of the suppression as the system size and the energy density reduces
 from the 50-70\% to the 70-90\% (70-80\%) centrality bin.

We note that our model predictions of minimum bias inclusive nuclear modification factors in OO collisions addresses the same $\langle N_{\rm part}\rangle \approx  10$ range
as 70-90\% (70-80\%) peripheral PbPb (XeXe) collisions.
Measuring $R_\text{AA}^h$ in OO collisions is a much wanted independent test of the expected system size dependence of parton energy loss,
that is free of assumptions about the modeling of the soft physics that enter the 
baseline of peripheral $R_\text{AA}^h$ measurements. We scrutinize the potential of discovering energy loss in small systems in our companion paper~\cite{Huss:2020dwe}.

\subsection{Robustness of model predictions}
\label{sec:robustness}

In the previous section we showed the results of a simple energy loss model based on the
BDMPS-Z energy loss formula of hard partons in a kinetically evolved background. The system size dependence was modeled by TrENTo initial state model and we included nPDF and fragmentation effects. Although this setup is well motivated, many of the model assumptions have not been independently constrained. Therefore we now stress-test the robustness of model predictions by varying  different model assumptions  in the same framework.

\subsubsection{Summary of models considered}
\label{sec:robustnessA}

First, to understand the relative importance of nPDF, system size modeling and fragmentation effects on our results, we consider four unphysical setups with some of these model components switched off. 
\begin{description}
\item[Minimal] In this minimal implementation, the isotropic background geometry is scaled according to optical Glauber $R$ and $\left<N_\text{part}\right>$. Energy loss is modeled only for gluons and no nPDF or fragmentation effects are included.  In essence, the gluon quenching factor \Eq{Qgp} with $n_g=6$ is used as a proxy for $R^h_\text{AA}$.
\item[Anisotropic] The same as \textbf{Minimal}, but the system size dependence of $R$ and $\left<N_\text{part}\right>$ is now modeled using the TrENTo initial state model and we include the average elliptic deformation of the background. 

\item[nPDF] The same as \textbf{Minimal}, but nPDF effects are included. That is the (partonic) gluon quenching factor is multiplied by ``vacuum" (partonic) $R^{\text{vac}}_\text{AA}$
shown in \Fig{fig:RAAnpdf}. The centrality dependence of nPDF effects is scaled with 
$\propto \left<N_{\rm part}-2\right>^{1/4}$, see \Eq{eq:RAAcent}. No fragmentation is included.

\item[Fragmentation]  The same as \textbf{Minimal}, but gluons are fragmented into hadrons, i.e., the hadronic quenching  factor \Eq{eq:Qzn} is compared to $R_\text{AA}^h$. No nPDF effects are included.
\end{description}

Next we study how our results depend on the assumed background medium evolution. As explained in \Sec{sec:geometry}, by default we use a particular simple parametrized temperature profile. Here, we test to what extent our predictions depend on this evolution. In all cases we include both  nPDF and fragmentation effects.
\begin{description}
\item[Simple] This is our default model described in \Sec{sec2} and with the results shown in \Sec{sec3a}. It includes geometry scaling based on TrENTo, nPDF effects, fragmentation and energy loss for both quarks and gluons.
\item[Simple $\mathbf{\tau_0 = 0.5}\,\text{fm}\mathbf{/c}$] The same as \textbf{Simple}, but the energy loss is calculated from the later starting time of $\tau_0 = 0.5\,\text{fm}/c$ instead of $\tau_0 = 0.05\,\text{fm}/c$.
\item [Simple $\mathbf{T_{\textbf{F}} = 120\,\text{MeV}}$] The same as \textbf{Simple}, but energy loss is computed up to a later time, namely when the temperature falls below $T_{\rm F}\,=120\,\text{MeV}$ instead of $T_{\rm F}\,=175\,\text{MeV}$. 
\item[Lattice EOS] The same as \textbf{Simple}, but the 
temperature profile is determined using lattice equation of state $T_\text{lat}(e)$~\cite{Bazavov:2014pvz}, where $e\approx 15T^4$ is the energy density in our (conformal) kinetic simulation.
The freeze-out temperature is again set to $T_{\rm F}\,=120\,\text{MeV}$.
\item[Bjorken] The same as \textbf{Simple}, but the kinetic temperature evolution  \Eq{Tevolution} is replaced by Bjorken scaling $T=T(\tau_i,\vec x) (\tau/\tau_i)^{-1/3}$ with $\tau_i=1\,\text{fm}/c$.
\item[Free streaming] The same as \textbf{Simple}, but with the free streaming ($\hat{\gamma}=0$) solution of kinetic theory for an azimuthally symmetric initial profile.
\end{description}

All model variations above used the parton energy loss formula derived by Arnold~\cite{Arnold:2008iy} in BDMPS-Z formalism. Here we use  our simple framework to compare three characteristically different parametrizations of parton energy loss inspired by recent phenomenological  studies~\cite{Djordjevic:2015hra,Zigic:2018smz,Noronha-Hostler:2016eow,Casalderrey-Solana:2014bpa}.
We calculate the shift function $S_{g/q}$ for these parametrizations with free normalization constant $\kappa$. 
\begin{description}
\item[A] Energy loss with weak path length and temperature dependence $dE/dL\propto -L^{0.4} T^{1.2}$, leading to 
\begin{equation}
    S_s = C_s \int_{\tau_0}^\infty d\tau \frac{\kappa}{6}\tau^{0.4} \,T(\tau, \vec{x}(\tau))^{1.2}.
\end{equation}
\item[B] Energy loss with linear path length dependence and strong temperature dependence, $dE/dL\sim -L T^3$, leading to
\begin{equation}
    S_s = C_s \int_{\tau_0}^\infty d\tau \frac{\kappa}{3}\tau \,T(\tau, \vec{x}(\tau))^3.
\end{equation}
\item[C] Energy loss
implementing stopping  with $dE/dL \propto -E_{\rm in} L^2/(L_\text{stop}^2\sqrt{L_\text{stop}^2-L^2})$
\begin{equation}
    S_s = C_s \int_{\tau_0}^\infty d\tau p_\perp \frac{4 \tau^2}{\pi \tau_{\rm stop}^2\sqrt{\tau_{\rm stop}^2 - \tau^2}},
\end{equation}
where $\tau_{\rm stop} = \frac{1}{2 (\kappa/5)} p_{\perp,0}^{1/3} T(\tau, \vec{x}(\tau))^{-4/3}$.
\end{description}

\subsubsection{Discussion}
\label{sec:robustnessB}

 \begin{figure}
    \centering
    \includegraphics[width=\linewidth]{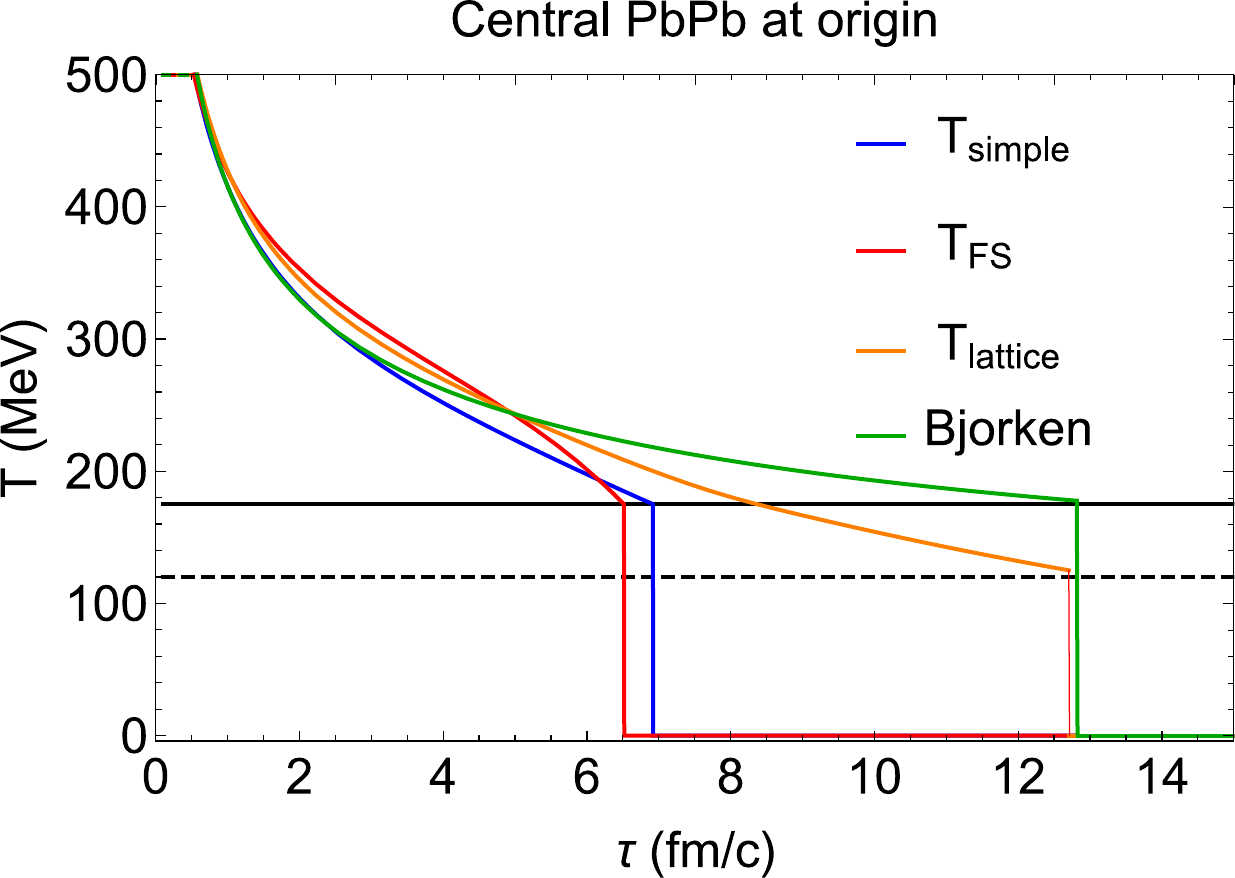}
    \caption{Comparison of time dependence of the temperature profile at the origin of a central PbPb collision in different considered temperature evolution scenarios. The freeze-out times ordered from shortest to longest are: free streaming (FS), Simple, Lattice and finally Bjorken expanding.  }
    \label{fig:hydrotemp}
\end{figure}

\begin{table*}
\caption{Values for the free model parameter $d=\hat{\bar{q}}/T^3$ or $\kappa$ extracted from the minimum bias point at $p_\perp = 54.4\,\text{GeV}$ in $\sqrt{s_{NN}}=5.02\,\text{TeV}$ PbPb collisions (see \Fig{fig:MinBiasPlot}).\label{tab:qhat}}
\begin{tabular}{lccccccc}
\hline
\hline
model  & nPDF & $\left<R\right>$, $\left<N_\text{part}\right>$,  $\left<N_\text{coll}\right>$& $\left<\epsilon_2\right>$ & $T$ evolution & Energy loss & Fragmentation & $\hat{\bar q}/T^3$ or $\kappa$ \\
\hline
 Minimal & no & optical Glauber & no & kinetic & BDMPS-Z & no & 0.89 $\pm $ 0.26 \\
 Anisotropic  & no & TrENTo& yes & kinetic & BDMPS-Z & no & 0.85 $\pm $ 0.24 \\
 nPDF  & yes & optical Glauber& no & kinetic & BDMPS-Z & no & 1.08 $\pm $ 0.27 \\
 Fragmentation  & no & optical Glauber& no & kinetic & BDMPS-Z & yes & 3.5 $\pm $ 1.1 \\
 Simple & yes & TrENTo& yes & kinetic & BDMPS-Z & yes & 4.3 $\pm $ 1.1 \\
 Simple, $\tau_0= 0.5 \text{fm}/c$  & yes & TrENTo & yes& kinetic & BDMPS-Z & yes & 8.1 $\pm $ 2.8 \\
 Simple, $T_\text{F} = 0.12 \text{GeV}$  & yes & TrENTo & yes& kinetic & BDMPS-Z & yes & 3.8 $\pm $ 0.9 \\
 Free streaming  & yes & TrENTo& no & free streaming & BDMPS-Z & yes & 2.69 $\pm $ 0.70 \\
 Lattice EOS & yes & TrENTo & yes& kinetic & BDMPS-Z & yes & 2.84 $\pm $ 0.70 \\
 Bjorken & yes & TrENTo & yes& $\propto \tau^{-1/3}$ & BDMPS-Z & yes & 3.59 $\pm $ 0.91 \\
 A  & yes & TrENTo& yes & kinetic & $dE/dx \propto \tau^{0.4} T^{1.2}$ & yes & 3.40 $\pm $ 0.71 \\
 B  & yes & TrENTo& yes & kinetic & $dE/dx \propto \tau T^3$ & yes & 4.32 $\pm $ 0.95 \\
 C  & yes & TrENTo & yes& kinetic & Stopping & yes & 2.54 $\pm $ 0.17 \\
 \hline
\hline
\end{tabular}
\end{table*}

\begin{figure*}
    \centering
    \includegraphics[width=\linewidth]{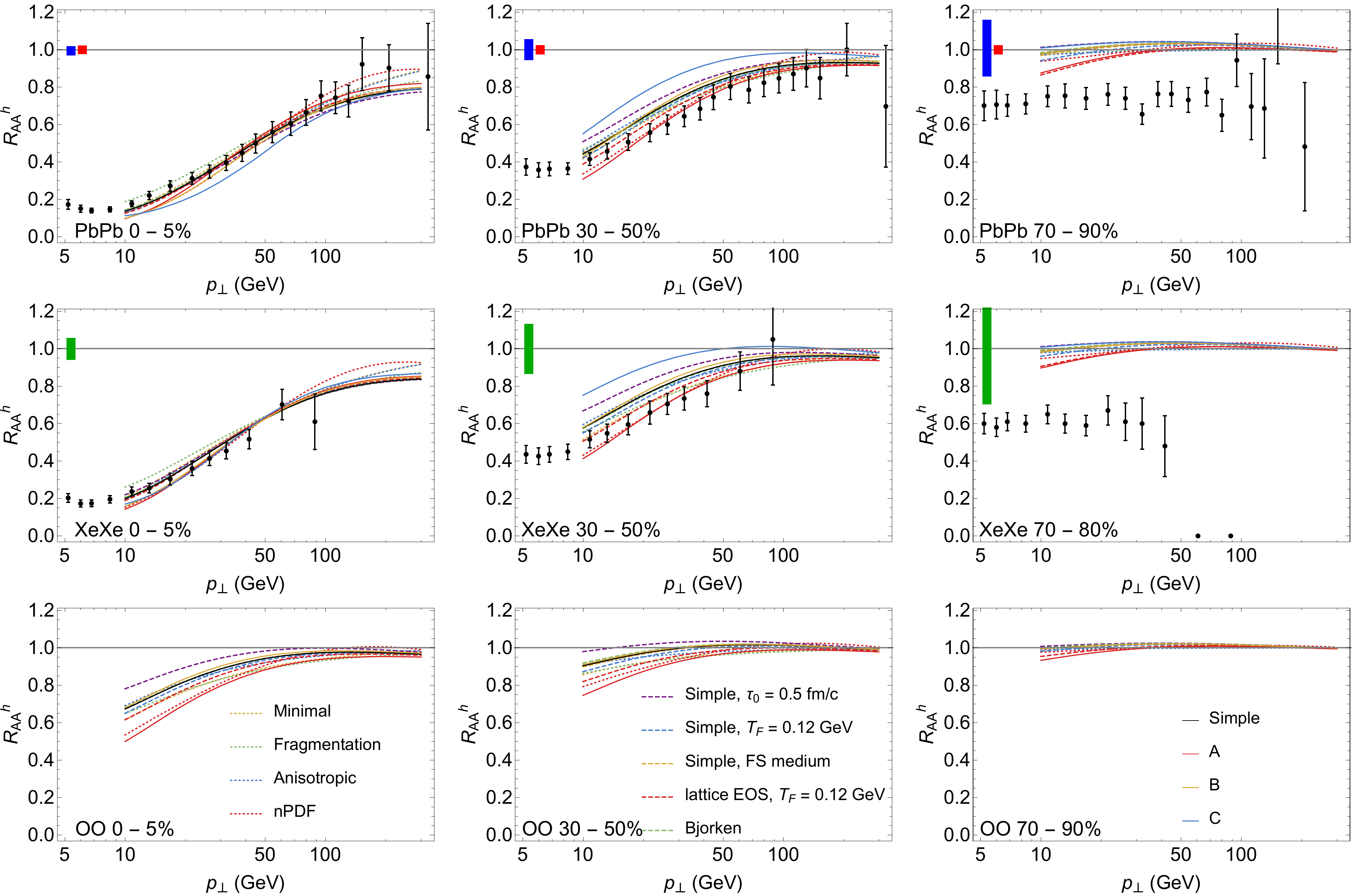}
    \caption{Charged hadron nuclear modification factors for different model scenarios in Table~\ref{tab:qhat}
    for (top) PbPb, (middle) XeXe and (bottom) OO  collisions in three centrality classes. Data points for PbPb and XeXe  are the same as in \Figs{fig:centralityPbPb} and \ref{fig:centralityXeXe}.
    }\label{fig:raamodels}
\end{figure*}

In Table~\ref{tab:qhat} we summarize the model variations introduced above. For each model we 
adjust the free parameter $d=\hat{\bar{q}}/T^3$ or $\kappa$ to reproduce the centrality averaged $R_\text{AA}^h$ at $p_\perp=54.4\,\text{GeV}$  in $\sqrt{s_{{NN}}} = 5.02\,\text{TeV}$ PbPb collisions 
as it was done in \Sec{sec3a}. We then compare these models in Fig.~\ref{fig:raamodels} to the centrality dependence of charged hadron nuclear modification factors measured in PbPb and XeXe collisions, and we extrapolate to  OO collisions at $\sqrt{s_{NN}} = 5.02\,\text{TeV}$.

Before entering a more detailed discussion,
let us note that despite the dramatic approximations implemented in the different models in Table~\ref{tab:qhat},
most of the models reproduce the $p_\perp$ dependence
of $R_{\text{AA}}^h$ in central and semi-central PbPb and XeXe collisions. They
do so with values of $\hat{\bar{q}}/T^3$ or $\kappa$ that vary significantly with model assumptions. However the aim of the present paper is only to estimate the expected signal of parton energy loss in light-ion collisions. We can do this extrapolation without judging the completeness of the different model scenarios or the numerical value of the extracted medium parameter  $\hat{\bar{q}}/T^3$ or $\kappa$.

Now we discuss individual model variations listed in   Table~\ref{tab:qhat}.
In the first four, \textbf{Minimal}, \textbf{Anisotropic}, \textbf{nPDF} and \textbf{Fragmentation}, some of the model components were switched off. The spread  of model predictions in \Fig{fig:raamodels} (dotted lines) informs us to what extent the detailed modeling of these components is important for system size extrapolations. 
Moreover, as fragmentation converts partons to much softer hadrons, the same $R_\text{AA}^h$ is achieved with three times larger value  of $\hat{\bar{q}}/T^3$. We note in addition that doing fragmentation directly of the quenched parton spectra in \Eq{Qh} instead of using \Eq{zn} increases $\hat{\bar{q}}/T^3$ by  $\approx 20\%$.

Next we considered the parton energy loss dependence on the variations of the background temperature evolution (dashed lines). 
Starting energy loss at $\mathbf{0.5\,\textbf{fm}/c}$ requires a twice larger value of $\hat{\bar{q}}/T^3$ than any other model variations in Table~\ref{tab:qhat}.
This model scenario shows also a more pronounced tension with experimental data in the mid-central PbPb and XeXe data. This suggests that data favors early onset of energy loss. Other variations of the temperature evolution---such as varying from \textbf{Bjorken} to \textbf{free-streaming}, extending the interaction down to $\mathbf{T_\textbf{F}=120\,\textbf{MeV}}$, or switching to \textbf{lattice EOS}, see  \Fig{fig:hydrotemp}---seem to have only a mild effect on $R_{\text{AA}}^h$.

We finally consider parton energy loss formulas that differ significantly from BDMPS-Z (solid lines). Here, the formula assuming full stopping (\textbf{C}) is arguably the most extreme choice, and it is the one that shows the most significant tension with the observed centrality dependence in PbPb and XeXe collisions. We therefore do not include it in our extrapolation to OO. The other two parametrizations (\textbf{A} and \textbf{B}) are comparable to our \textbf{Simple} model.

Given the range of model assumptions explored, we regard the envelope of the different
predictions in \Fig{fig:raamodels}  as a realistic theory uncertainty for $R_\text{AA}^h$ in OO collisions at  $\sqrt{s_\text{NN}}=5.02\,\text{TeV}$. In our companion paper~\cite{Huss:2020dwe}, we use the same range of model scenarios to compute the expected parton energy loss signal and its uncertainty for the proposed $\sqrt{s_\text{NN}}=7\,\text{TeV}$ OO collisions.

\begin{figure}
    \centering
    \includegraphics[width=\linewidth]{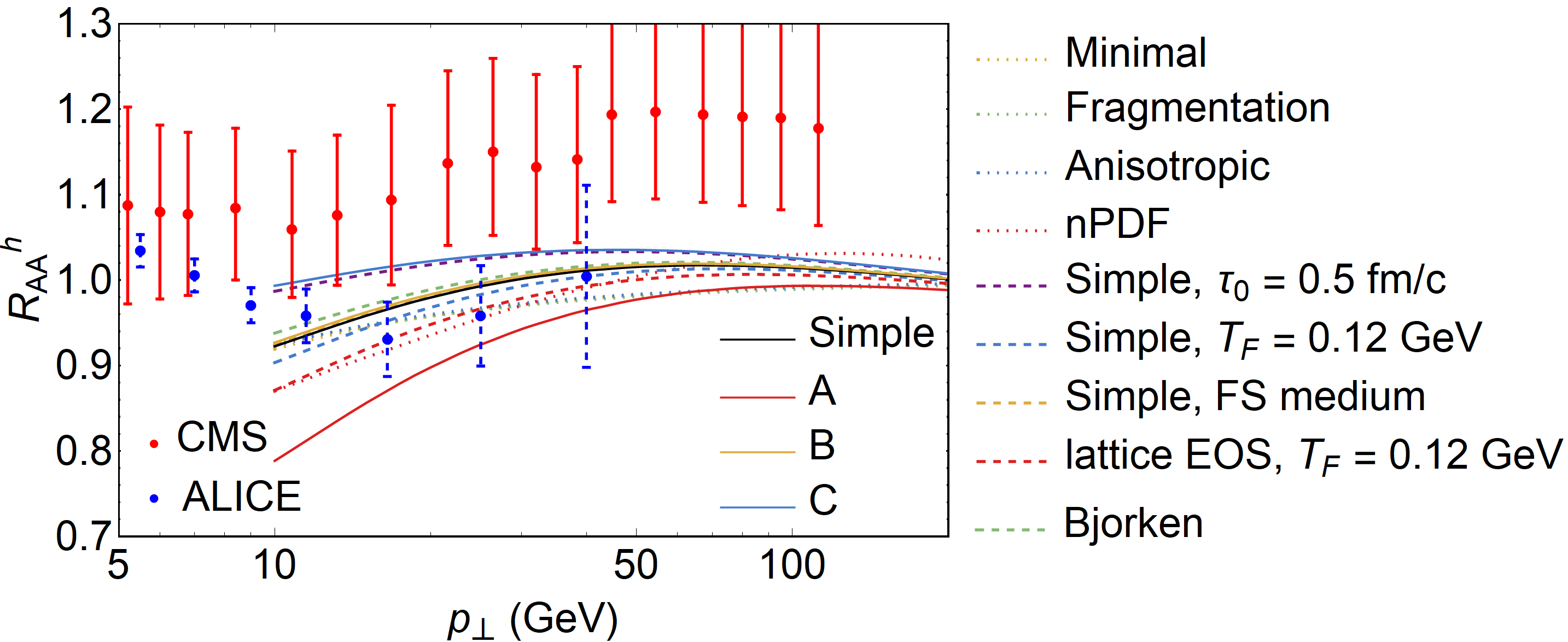}
    \caption{Charged hadron nuclear modification factors for different model scenarios in Table~\ref{tab:qhat}
    for minimum bias $p$Pb collisions. Data points are from \cite{Khachatryan:2016odn, Acharya:2018qsh}.}
    \label{fig:ppb}
\end{figure}
For completeness we show in \Fig{fig:ppb} results for the same set of model variations applied to minimum bias $p$Pb. As there are no mechanisms in the considered models (other than nPDF effects) to produce larger than unity nuclear modification, none of the models go through the experimental data points.

\subsection{High-momentum hadron anisotropy}
\label{sec:v2}

\begin{figure*}
    \centering
    \includegraphics[width=\linewidth]{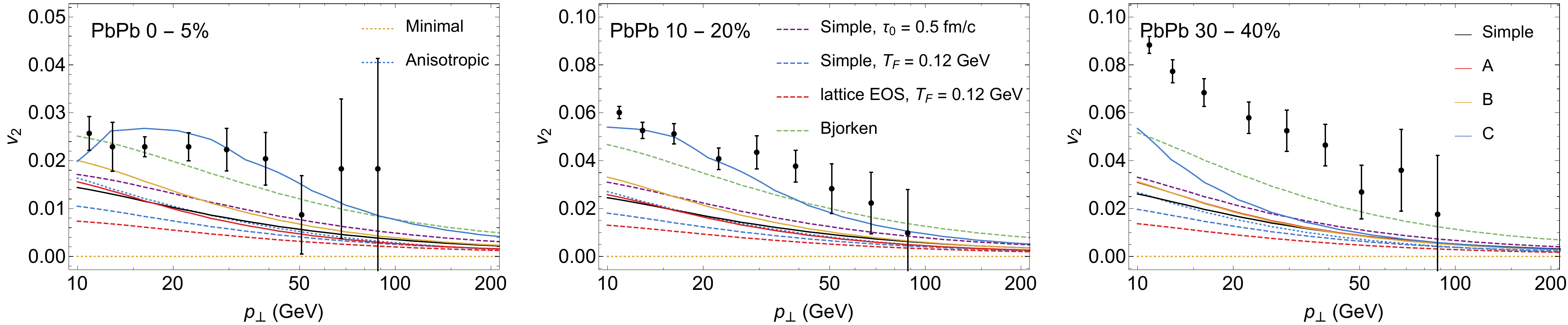}
    \caption{Centrality and $p_\perp$ dependence of elliptic flow coefficient $v_2$ for high-$p_\perp$ hadrons in different model scenarios, Table~\ref{tab:qhat}, in $\sqrt{s_{NN}}=5.02\,\text{TeV}$ PbPb collisions together with the experimental data~\cite{Sirunyan:2017pan}. The \textbf{Minimal} (and not shown \textbf{Fragmentation}, \textbf{nPDF}, and \textbf{Free streaming}) scenario does not implement initial deformation of the geometry and $v_2$ is zero.}
    \label{fig:v2models}
\end{figure*}

A more differential probe of parton energy loss is the high-momentum anisotropy of the final particles. In a peripheral collision with elliptical shape parton energy loss is expected to depend on the orientation of the hard parton trajectory.
This dependence can be parametrized as $\cos(2\phi)$ modulation of the nuclear modification factor
\begin{equation}
    R_\text{AA}^h(p_\perp,\phi)= R_\text{AA}^h(p_\perp)\left[ 1+2v_2(p_\perp) \cos (2\phi-2\phi_2)) \right],\label{eq:v2}
\end{equation}
where $\phi$ is the azimuthal momentum angle and $\phi_2$ characterizes the event-plane.
Experimentally,  $v_2(p_\perp)$ is obtained from the correlation between high-$p_\perp$ hadrons and soft particles.

It has been long a challenge to simultaneously describe 
the nuclear modification factor and the sizable high-momentum anisotropy
within the same model.
Models that do not include early time parton energy loss
typically fare better~\cite{Andres:2016iys}, because they concentrate the energy loss at later times where the background anisotropy is more relevant. Moreover, it has been shown that including event-by-event fluctuations of the underlying medium can increase the high-$p_T$ elliptic flow~\cite{Noronha-Hostler:2016eow}.

Our simple framework does not model event-by-event fluctuations of soft particle production and therefore we do not expect it to accurately reproduce the experimentally measured $v_2(p_\perp)$.
Nevertheless, it is interesting to check how different model assumptions in \Sec{sec:robustness} affect the elliptic flow of high-$p_\perp$. We determine $v_2(p_\perp)$ from the energy loss modulation of \Eq{eq:v2} with respect to the inputted background deformation. In \Fig{fig:v2models} we compare our model predictions of $v_2(p_\perp)$  to data in different centrality bins of $\sqrt{s_{NN}}=5.02\,\text{TeV}$
 PbPb collisions. We see that for most of the model scenarios $v_2(p_\perp)$ is underpredicted by a factor of $\approx 2$.
Possible exceptions are the scenario with Bjorken temperature profile and energy loss model with stopping \textbf{C}.
The slow temperature evolution in \textbf{Bjorken} and
the concentration of energy loss towards the end of the evolution in model \textbf{C}
presumably allow for stronger correlation between initial state geometry and high-$p_\perp$ energy loss.

\begin{figure*}
    \centering
    \includegraphics[width=0.49 \linewidth]{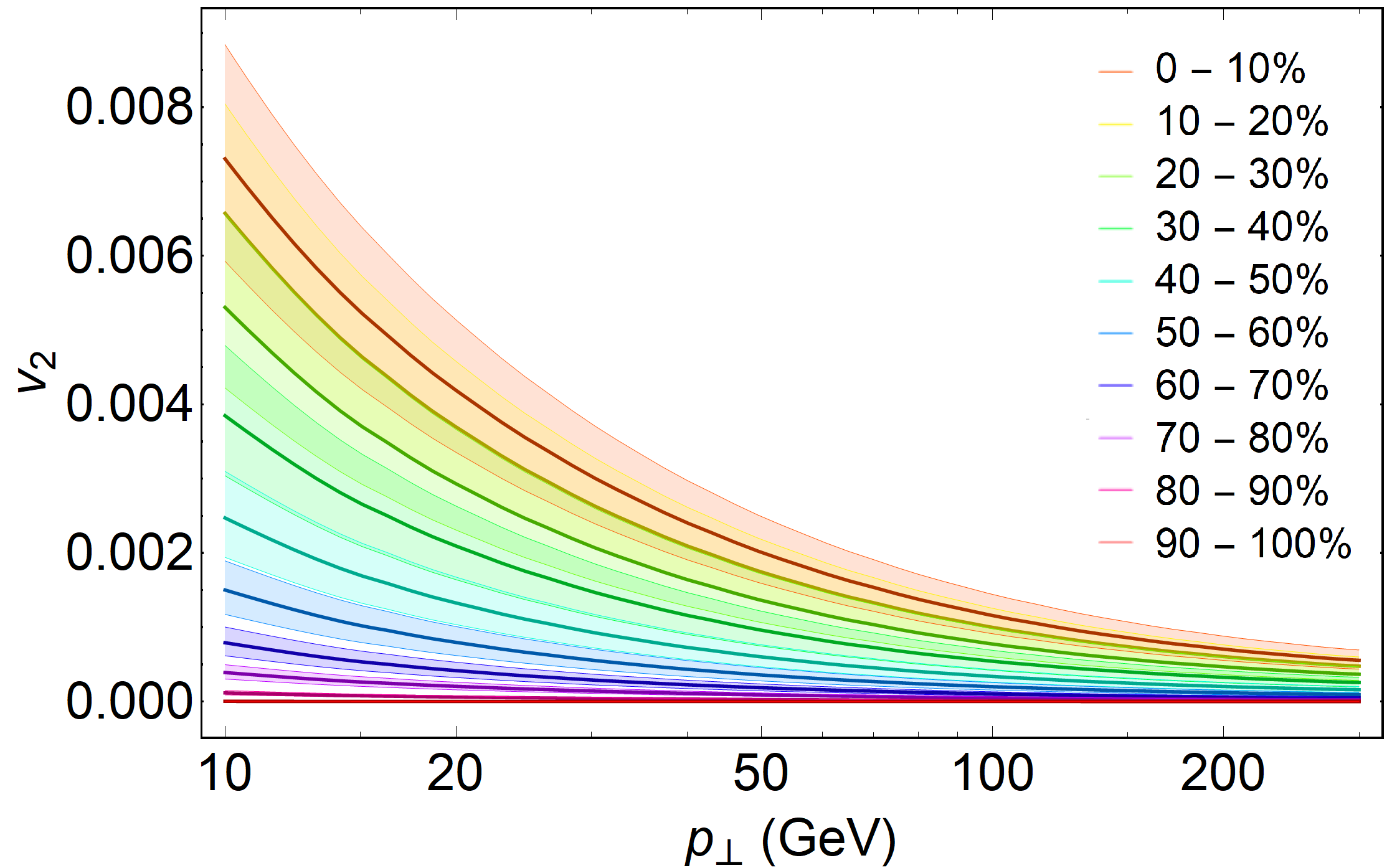}
     \includegraphics[width=0.49 \linewidth]{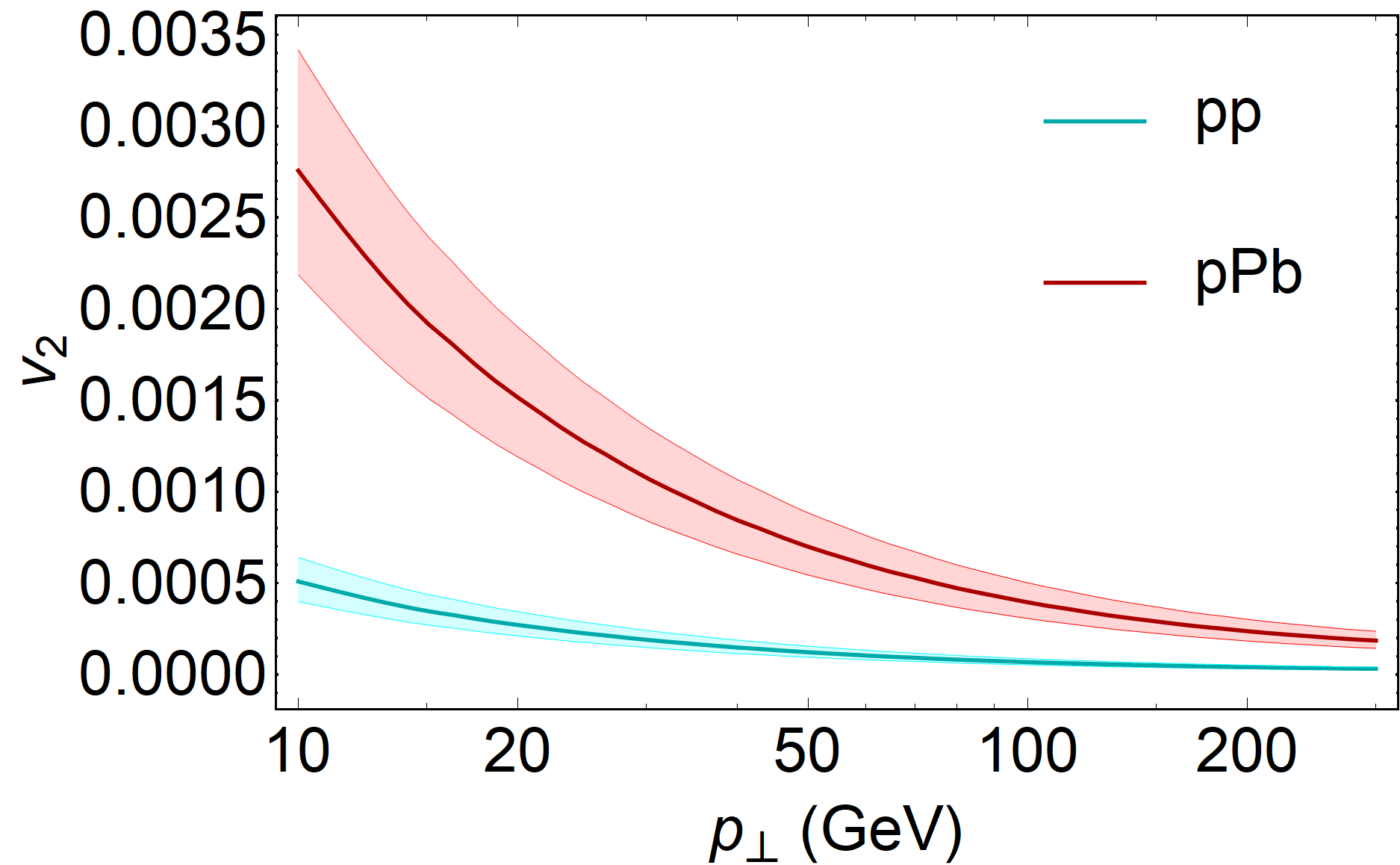}
    \caption{(left)  Elliptic flow coefficient $v_2$ in $\sqrt{s_{NN}}=5.02\,\text{TeV}$ OO collisions in several centrality classes for the \textbf{Simple} model (curves are ordered from high to low centrality). As opposed to PbPb collisions the $v_2$ is strongest in central collisions, as in these smaller systems fluctuations are more important than in PbPb collisions.
(right) Elliptic flow coefficient $v_2$ in minimum bias  $pp$ (lower band) and $p$Pb (upper band) collisions for the \textbf{Simple} model at $\sqrt{s_{NN}}=5.02\,\text{TeV}$.
    \label{fig:ppbv2}
}
    \label{fig:OOsimple}
\end{figure*}

Finally in \Fig{fig:ppbv2}
we show the \textbf{Simple} model predictions for $v_2(p_\perp)$ in small collisions systems, i.e., centrality selected OO collisions and minimum bias $pp$ and $p$Pb. The tendency of our model to underpredict the experimental data prevents us from making quantitative conclusions about high-$p_\perp$ elliptic flow in small systems. However we can make the following qualitative observations. First the large initial eccentricity in central OO collisions~\cite{Katz:2019qwv, Rybczynski:2019adt} results in monotonically decreasing $v_2(p_\perp)$ with centrality. 
Secondly, we find  a small elliptic flow in minimum bias $p$Pb  collisions and even smaller in $pp$.
Making more quantitative statements about the elliptic flow magnitude in small systems is outside the scope of the current paper.

\section{Conclusions}
\label{conclusions}

In the present paper we document a model for calculating the high-momentum charged hadron spectra modifications due to the medium induced parton energy loss in small collision systems.
Our baseline calculation of hadron spectra consists of the leading order QCD partonic cross sections convoluted with (nuclear modified) parton distribution functions and fragmentation functions. The parton energy loss is modeled by small-$x$ gluon emission 
and the dynamical temperature profile is scaled to match the expected system size and entropy.

After tuning a single model parameter to a single data point of the charged hadron nuclear modification factor at $p_\perp\approx 50\,\text{GeV}$ in minimum bias PbPb collisions, we demonstrated that our model is consistent---up to a 2-$\sigma$ tension in the most peripheral bin---with the $p_\perp$ and centrality dependence of $R^h_\text{AA}$ in $\sqrt{s_{NN}} = 5.02\,\text{TeV}$   PbPb and $\sqrt{s_{NN}} = 5.44\,\text{TeV}$ XeXe collisions.
Validated against these data, the model provides well motivated predictions for the charged hadron nuclear modification factors in the minimum bias $p$Pb, OO, and ArAr collisions and in centrality selected OO collisions.

To ascertain the systematic uncertainties we varied the different model components, medium evolution,  and energy loss formula. All modeling scenarios provide rather comparable momentum and system size dependencies of $R_\text{AA}^h$ once fitted to the same point in the minimum bias PbPb collisions. These model variations predict up to $\approx 15\%$ modification of  hadron spectra in minimum bias OO collisions at $p_\perp\approx 50\,\text{GeV}$. Such small nuclear modification could not be resolved within the systematic experimental uncertainties present in the comparable size peripheral PbPb or XeXe collisions.
However, a measurement of $R^h_\text{AA}$ in an inclusive OO collisions is free of  model dependent uncertainties entering the centrality selected nuclear modification factor.

The ability to identify parton energy loss also depends on the accuracy with which the baseline without the medium effects can be calculated. 
In the companion paper~\cite{Huss:2020dwe}, we show that the accuracy of the baseline $R_{\rm AA}^{h,{\rm vac}}$ in inclusive OO collisions is
known with sufficient precision that the discovery of the medium induced parton energy loss in small systems with $\langle N_{\rm part} \rangle \approx 10$ is possible.

{\bf Acknowledgements.}
We thank Liliana Apolinário, James Mulligan and Daniel Pablos for discussions.
We thank Hannu Paukkunen for providing EPPS16 parton distribution functions for oxygen and 
Guilherme Milhano for numerous discussions throughout the project.

\bibliography{master.bib}

\appendix
\section{Modeling collision geometry}
\label{sec:geom}

Parton energy loss is sensitive to the spatiotemporal extension of the QCD medium and its density profile. In the main text, we have described the physical picture underlying our modeling of the collision geometry. For completeness, we provide in this appendix quantitative information. 

The simplest way to determine the initial geometry for PbPb, XeXe and OO collisions is to use  the standard optical Glauber model~\cite{Miller:2007ri}. In Table~\ref{tableNc} we present, for each collision system, the computed number of participants $N_{\rm part}$, the number of binary collisions $N_{\rm coll}$, the radius of the profile $R$, and the rms radius $R_h$ as a function of centrality defined by the impact parameter $b$.
\begin{table}[ht]
\caption{The values describing collision geometry in PbPb, XeXe, and OO collisions taken from optical Glauber model with $\sigma_{NN} = 64\,\text{mb}$. The two parameter Fermi distribution was used for the nuclear geometry of PbPb and XeXe, while for OO the three parameter Fermi distribution was used.}
\begin{center}
\begin{tabular}{cccccc}
\hline
\hline
PbPb, centrality & b [fm] & $N_{\rm part}$ & $N_{\rm coll}$ & $R$  [fm] & $R_{h}$  [fm] \\
\hline
 0.025 & 2.45919 & 375.375 & 1763.98 & 4.20174 & 3.60552 \\
 0.075 & 4.25642 & 321.194 & 1363.77 & 3.91481 & 3.34661 \\
 0.2 & 6.95066 & 203.666 & 720.981 & 3.4475 & 2.88248 \\
 0.4 & 9.82971 & 90.7868 & 219.436 & 2.83458 & 2.37449 \\
 0.6 & 12.0389 & 30.0211 & 44.2891 & 2.38618 & 2.08597 \\
 0.8 & 13.9018 & 6.00284 & 5.57849 & 2.16284 & 2.04995 \\
\end{tabular}

\begin{tabular}{cccccc}
XeXe, centrality & b[fm] & $N_{\rm part}$ & $N_{\rm coll}$ & $R$ [fm] & $R_{h}$ [fm] \\
\hline
 0.025 & 2.11583 & 232.567 & 908.333 & 3.59924 & 3.09502 \\
 0.075 & 3.66213 & 196.754 & 704.134 & 3.41962 & 2.89356 \\
 0.2 & 5.98019 & 129.546 & 372.711 & 2.99913 & 2.52769 \\
 0.4 & 8.45726 & 57.6395 & 115.241 & 2.51918 & 2.14456 \\
 0.6 & 10.358 & 19.7876 & 25.5844 & 2.19453 & 1.95756 \\
 0.8 & 11.9629 & 4.67246 & 4.12596 & 2.0568 & 1.96054 \\
\end{tabular}

\begin{tabular}{cccccc}
OO, centrality & b[fm] & $N_{\rm part}$ & $N_{\rm coll}$ & $R$ [fm] & $R_{h}$ [fm] \\
\hline
 0.025 & 1.04907 & 25.5465 & 45.5075 & 1.94378 & 1.66221 \\
 0.075 & 1.81576 & 22.9137 & 37.1249 & 1.87315 & 1.62262 \\
 0.2 & 2.9651 & 15.9454 & 22.1546 & 1.77075 & 1.54863 \\
 0.4 & 4.1933 & 8.44554 & 9.29498 & 1.63558 & 1.48786 \\
 0.6 & 5.14062 & 4.05636 & 3.67912 & 1.57844 & 1.48136 \\
 0.8 & 6.04545 & 1.61284 & 1.23164 & 1.5594 & 1.52022 \\
 \hline
\hline
\end{tabular}
\label{tableNc}
\end{center}
\end{table}

The more sophisticated way to determine the initial geometry for each collision system is to use the TrENTo initial condition framework \cite{Moreland:2014oya}. In the TrENTo model, the initial transverse entropy density profile is computed from
\begin{equation}
s(x,y) \propto \left (\frac{T^p_A + T^p_B}{2} \right )^{1/p},  \label{eq:entr} 
\end{equation}
where the parameter $p$ controls the mixing of fluctuating thickness functions $T_A$ and $T_B$. In this paper, we use the following parameter values \cite{Moreland:2018gsh} to obtain the entropy density for each collision system: 
\begin{itemize}
\item  reduced thickness parameter $p=0.013$
\item fluctuation parameter $k=0.93$
\item nucleon width $\sigma =0.6$
\item inelastic nucleon-nucleon cross section $\sigma_{NN} = 64~\text{mb}$
\end{itemize}

For all elements we used the standard settings in TrENTo, except for oxygen where for the nucleon positions we used the tables from \cite{Lonardoni:2017egu} (see also Ref.!\cite{Rybczynski:2019adt}), as provided in Ref.~\cite{Alver:2008aq}.

We take an ensemble of 20 000 events and for each centrality (defined as a class of events $\pm 5\%$ from the midpoint value) we obtain an average of all values used in the main text, which is the radius of the entropy density (Table~\ref{TrENToradius}), the average entropy density (Table~\ref{TrENToenergydensity}, used in \Eq{eq:tempvsS}), the radius of the hard parton scattering centers $R_h$ (Table~\ref{TrENToncoll}), the number of participating nucleons $N_{\rm part}$ (Table~\ref{TrENTonpart}, used for nPDF corrections), the eccentricity $\epsilon_2$ (Table~\ref{TrENToeps2}) and finally $N_{\rm coll}$ (Table~\ref{TrENToncollw}, used for weighting centrality classes).

\begin{table}
\begin{center}
\begin{tabular}{ccccccc}
 \text{centrality} & \text{PbPb} & \text{XeXe} & \text{OO} & \text{ArAr} & \text{pp} & \text{pPb} \\
\hline
 0.05 & 4.274 & 3.659 & 2.153 & 2.762 & 1.08 & 1.431 \\
 0.15 & 3.847 & 3.31 & 2.026 & 2.561 & 1.087 & 1.416 \\
 0.25 & 3.504 & 3.036 & 1.904 & 2.394 & 1.083 & 1.412 \\
 0.35 & 3.205 & 2.804 & 1.787 & 2.242 & 1.082 & 1.383 \\
 0.45 & 2.937 & 2.572 & 1.663 & 2.093 & 1.059 & 1.35 \\
 0.55 & 2.701 & 2.375 & 1.521 & 1.913 & 1.06 & 1.287 \\
 0.65 & 2.444 & 2.132 & 1.361 & 1.684 & 1.034 & 1.242 \\
 0.75 & 2.148 & 1.827 & 1.187 & 1.432 & 1.013 & 1.149 \\
 0.85 & 1.813 & 1.392 & 1.03 & 1.144 & 0.984 & 1.05 \\
 0.95 & 1.503 & 0.975 & 0.899 & 0.926 & 0.925 & 0.921 \\
 \end{tabular}
\caption{The rms radius $R$ of entropy density \Eq{eq:entr} as a function of centrality for different collision systems (in fm).}
\label{TrENToradius}
\end{center}
\end{table}

\begin{table}
\caption{ The average entropy density per unit rapidity (in arbitrary units), as defined by $dS/dy/\pi R^2$, with $dS/dy$ the transverse integral of \Eq{eq:entr} and $R$ as given in Table~\ref{TrENToradius}.}
\label{TrENToenergydensity}
\begin{center}
\begin{tabular}{ccccccc}
\hline
\hline
 \text{centrality} & \text{PbPb} & \text{XeXe} & \text{OO} & \text{ArAr} & $pp$ & $p$Pb \\
\hline
 0.05 & 2.782 & 2.323 & 0.746 & 1.197 & 0.26 & 0.457 \\
 0.15 & 2.281 & 1.848 & 0.587 & 0.938 & 0.191 & 0.354 \\
 0.25 & 1.833 & 1.442 & 0.467 & 0.723 & 0.162 & 0.295 \\
 0.35 & 1.394 & 1.063 & 0.366 & 0.529 & 0.14 & 0.259 \\
 0.45 & 0.995 & 0.751 & 0.284 & 0.378 & 0.126 & 0.226 \\
 0.55 & 0.66 & 0.492 & 0.219 & 0.269 & 0.107 & 0.197 \\
 0.65 & 0.409 & 0.303 & 0.167 & 0.193 & 0.093 & 0.159 \\
 0.75 & 0.229 & 0.182 & 0.127 & 0.142 & 0.076 & 0.129 \\
 0.85 & 0.108 & 0.116 & 0.086 & 0.105 & 0.057 & 0.094 \\
 0.95 & 0.028 & 0.057 & 0.038 & 0.048 & 0.031 & 0.046 \\
 \hline\hline
 \end{tabular}
\end{center}
\end{table}

\begin{table}
\caption{
RMS radius of the hard parton scattering centers, $R_h$ (in fm). TrENTo does not directly output the hard parton density, but since for $p\approx 0$ the entropy density is obtained from the thickness functions $T_A$ and $T_B$ by $\propto \sqrt{T_A T_B}$, a good proxy for the hard scattering center density $n_\text{coll}\propto T_A T_B$ can be found by squaring the entropy density.}
\label{TrENToncoll}
\begin{center}
\begin{tabular}{ccccccc}
\hline
\hline
 \text{centrality} & \text{PbPb} & \text{XeXe} & \text{OO} & \text{ArAr} & $pp$ & $p$Pb \\
\hline
 0.05 & 3.541 & 3.005 & 1.732 & 2.232 & 0.821 & 1.138 \\
 0.15 & 3.131 & 2.695 & 1.632 & 2.068 & 0.827 & 1.124 \\
 0.25 & 2.802 & 2.457 & 1.534 & 1.939 & 0.825 & 1.122 \\
 0.35 & 2.53 & 2.274 & 1.446 & 1.83 & 0.824 & 1.099 \\
 0.45 & 2.294 & 2.093 & 1.346 & 1.722 & 0.796 & 1.064 \\
 0.55 & 2.097 & 1.957 & 1.221 & 1.582 & 0.794 & 1.006 \\
 0.65 & 1.874 & 1.769 & 1.074 & 1.375 & 0.772 & 0.964 \\
 0.75 & 1.618 & 1.508 & 0.913 & 1.141 & 0.753 & 0.879 \\
 0.85 & 1.34 & 1.1 & 0.765 & 0.871 & 0.728 & 0.788 \\
 0.95 & 1.09 & 0.718 & 0.648 & 0.673 & 0.671 & 0.669 \\
 \hline
\hline
 \end{tabular}
\end{center}
\end{table}
\begin{table}
\caption{ The number of participating nucleon collisions  $N_{\rm part}$ for several collision systems and centrality classes.} \label{TrENTonpart}
\begin{center}
\begin{tabular}{ccccccc}
\hline
\hline
 \text{centrality} & \text{PbPb} & \text{XeXe} & \text{OO} & \text{ArAr} & $pp$ & $p$Pb \\
\hline
 0.05 & 362.391 & 226.252 & 26.191 & 67.6 & 2.0 & 13.457 \\
 0.15 & 270.507 & 168.425 & 20.826 & 51.677 & 2.0 & 11.055 \\
 0.25 & 193.056 & 120.272 & 16.09 & 37.977 & 2.0 & 9.604 \\
 0.35 & 131.471 & 81.684 & 12.228 & 26.676 & 2.0 & 8.085 \\
 0.45 & 83.977 & 52.408 & 9.13 & 18.28 & 2.0 & 6.724 \\
 0.55 & 50.561 & 31.788 & 6.564 & 12.01 & 2.0 & 5.049 \\
 0.65 & 28.437 & 17.662 & 4.676 & 7.498 & 2.0 & 3.894 \\
 0.75 & 14.105 & 9.095 & 3.292 & 4.7 & 2.0 & 2.921 \\
 0.85 & 6.18 & 4.313 & 2.486 & 2.952 & 2.0 & 2.402 \\
 0.95 & 2.752 & 2.332 & 2.105 & 2.168 & 2.0 & 2.116 \\
 \hline
\hline
 \end{tabular}
\end{center}
\end{table}

\begin{table}
\caption{ The average ellipticity $\epsilon_2$, defined as the ratio of the entropy weighted averages $\langle (x^2+y^2)e^{2i\arctan(y/x)}\rangle/\langle (x^2+y^2)\rangle$.} \label{TrENToeps2}
\begin{center}
\begin{tabular}{ccccccc}
\hline
\hline
 \text{centrality} & \text{PbPb} & \text{XeXe} & \text{OO} & \text{ArAr} & $pp$ & $p$Pb \\
\hline
 0.05 & 0.124 & 0.111 & 0.206 & 0.155 & 0.325 & 0.337 \\
 0.15 & 0.231 & 0.211 & 0.257 & 0.216 & 0.328 & 0.34 \\
 0.25 & 0.305 & 0.281 & 0.295 & 0.264 & 0.32 & 0.349 \\
 0.35 & 0.352 & 0.339 & 0.33 & 0.321 & 0.332 & 0.342 \\
 0.45 & 0.391 & 0.375 & 0.364 & 0.372 & 0.317 & 0.351 \\
 0.55 & 0.413 & 0.414 & 0.38 & 0.41 & 0.317 & 0.343 \\
 0.65 & 0.416 & 0.449 & 0.392 & 0.439 & 0.312 & 0.359 \\
 0.75 & 0.366 & 0.47 & 0.371 & 0.429 & 0.315 & 0.35 \\
 0.85 & 0.266 & 0.432 & 0.324 & 0.357 & 0.309 & 0.328 \\
 0.95 & 0.093 & 0.297 & 0.262 & 0.275 & 0.276 & 0.277 \\
 \hline
\hline
 \end{tabular}
\end{center}
\end{table}

\begin{table}
\caption{ The number of binary collisions  $N_{\rm coll}$ for several collision systems and centrality classes.} \label{TrENToncollw}
\begin{center}
\begin{tabular}{ccccccc}
\hline
\hline
 \text{centrality} & \text{PbPb} & \text{XeXe} & \text{OO} & \text{ArAr} & $pp$ & $p$Pb \\
\hline
 0.05 & 1358.26 & 718.43 & 32.498 & 121.841 & 1.0 & 12.457 \\
 0.15 & 810.928 & 412.007 & 22.298 & 75.182 & 1.0 & 10.055 \\
 0.25 & 468.292 & 235.183 & 15.196 & 46.276 & 1.0 & 8.604 \\
 0.35 & 254.072 & 126.544 & 10.319 & 27.488 & 1.0 & 7.085 \\
 0.45 & 126.224 & 64.936 & 6.99 & 16.211 & 1.0 & 5.724 \\
 0.55 & 59.738 & 31.91 & 4.578 & 9.481 & 1.0 & 4.049 \\
 0.65 & 26.932 & 14.715 & 3.024 & 5.284 & 1.0 & 2.894 \\
 0.75 & 11.019 & 6.515 & 1.958 & 3.008 & 1.0 & 1.921 \\
 0.85 & 4.066 & 2.663 & 1.363 & 1.709 & 1.0 & 1.402 \\
 0.95 & 1.529 & 1.238 & 1.082 & 1.128 & 1.0 & 1.116 \\
 \hline
\hline
 \end{tabular}
\end{center}
\end{table}

\section{Parton and hadron production\label{sec:cross}}

In this section we summarize the LO computations of inclusive parton and hadron cross sections including the discussion of parton distribution and fragmentation functions.

\subsection{Single inclusive parton cross section}

At LO in the strong coupling $\alpha_s$, the production of jets in the collisions of hadrons $A$ and $B$ with momentum $P_A$ and $P_B$ is given by the partonic $2\to 2$ QCD scattering process
\begin{equation}
a(p_a) + b(p_b) \to c(p_1)  +  d(p_2),
\end{equation}
where two incoming partons $a$ and $b$ are sampled from parton distribution functions and the scattered partons $c$ and $d$ can be identified as final state jets.

According to the factorization theorem, the total two-jet cross section may be written as
\begin{equation}
\label{eq:ABtojet}
\sigma^{AB}_{cd} = \sum_{a,b} \int {d}x_A {d}x_B f^A_{a}(x_A,\mu_F^2)f^B_{b}(x_B,\mu_F^2) \, \hat \sigma^{ab}_{cd},
\end{equation}
where the partonic $ab\to cd$ cross section $\hat{\sigma}_{cd}^{ab}$ is convoluted with parton distribution functions $f^{A}_{a}$ and $f^{B}_{b}$ (evaluated at factorization scale $\mu_F$) describing the number density of finding a parton with a given momentum fraction $x_{A}$ and $x_B$ inside the hadron, i.e. $p_a =x_A P_A$ and $p_b = x_BP_B$. In this paper, the CT14 parametrization \cite{Dulat:2015mca} is used as the $pp$ baseline PDFs, and the nuclear modifications are taken from the EPPS16 \cite{Eskola:2016oht} for the O and Pb nucleus. 

In the hadronic center-of-mass frame, the four-momenta of the incoming partons, in the light-cone coordinates $(+,-,\perp)$, can be expressed in terms of the momentum fraction variables $x_A$ and $x_B$ as:
\begin{equation}
p_a = x_A\sqrt{\frac{s}{2}}(1,0,\mathbf{0}_{\perp}), \quad p_b = x_B\sqrt{\frac{s}{2}}(0,1,\mathbf{0}_{\perp}),
\end{equation}
where $s \equiv (P_A + P_B)^2$  denotes the center-of-mass energy squared. The jet four-momenta $p_1$  and $p_2$ can be parametrized in terms of the transverse momentum $p_{\perp}$ and rapidities $y_1$ and $y_2$ as:
\begin{equation}
\begin{split}
p_1 &= (\frac{p_{\perp}}{\sqrt{2}}e^{y_1}, \frac{p_{\perp}}{\sqrt{2}}e^{-y_1}, \mathbf{p}_{\perp 1}), \\
p_2 &= (\frac{p_{\perp}}{\sqrt{2}}e^{y_2}, \frac{p_{\perp}}{\sqrt{2}}e^{-y_2}, \mathbf{p}_{\perp 2}),
\end{split}
\end{equation}
where $\mathbf{p}_{\perp 1} = - \mathbf{p}_{\perp 2}$ and $\vert\mathbf{p}_{\perp 1}\vert = \vert\mathbf{p}_{\perp 2} \vert \equiv p_{\perp}$. The momentum conservation fixes $x_A$ and $x_B$:
\begin{equation}
x_A = \frac{p_{\perp}}{\sqrt{s}}\left (e^{y_1} + e^{y_2}\right ), \quad x_B = \frac{p_{\perp}}{\sqrt{s}}\left (e^{-y_1} + e^{-y_2}\right ),
\end{equation}
where $0<x_{A/B}<1$.

The partonic cross section $\hat\sigma^{ab}_{cd}$ can be evaluated in perturbative QCD  by the standard formula
for massless partons
\begin{align}
  \hat\sigma^{ab}_{cd} &=  \frac{1}{1+\delta_{cd}}\frac{1}{4 p_a\cdot p_b}
  \int \frac{d^3p_1}{(2\pi)^3 2 E_{p_1}}
  \int \frac{d^3p_2}{(2\pi)^3 2 E_{p_2}}\nonumber\\
  \times &(2\pi)^{4}\delta^{(4)}(p_a+p_b-p_1-p_2) \langle \vert \mathcal{M}(ab \to cd)\vert^2 \rangle.
\end{align}
Here, the Lorentz invariant phase space elements are multiplied with $\langle \vert\mathcal{M}\vert^2 \rangle$, which is the invariant matrix element square averaged (summed) over initial (final) state spin/polarisation and color.
The partonic cross section depends on the partonic Mandelstam variables:
\begin{align}
\begin{split}
\hat s &\equiv (p_a + p_b)^2 =2p_{\perp}^2(1+\cosh(y_1-y_2)),\\
\hat t &\equiv (p_a  - p_1)^2=-p_{\perp}^2(1+e^{-(y_1-y_2)}), \\
\hat u &\equiv (p_b - p_1)^2= -p_{\perp}^2(1+e^{+(y_1-y_2)}).
\end{split}
\end{align}
Neglecting quark masses, there are only eight flavor independent $2 \to 2$ partonic processes at LO \cite{Ellis:1985er}:
\begin{subequations}
\begin{align}
\langle \vert\mathcal{\hat{M}}(qq' \to qq')\vert^2 \rangle
& = \frac{4}{9} \frac{\hat{s}^2+\hat{u}^2}{\hat{t}^2}, \\
\langle \vert\mathcal{\hat{M}}(qq \to qq)\vert^2 \rangle
& = \frac{4}{9} \left( \frac{\hat{s}^2+\hat{u}^2}{\hat{t}^2} + \frac{\hat{s}^2+\hat{t}^2}{\hat{u}^2} \right) -\frac{8}{27}\frac{\hat{s}^2}{\hat{t}\hat{u}}, \\
\langle \vert\mathcal{\hat{M}}(q \bar q \to q'\bar q')\vert^2 \rangle
& = \frac{4}{9} \frac{\hat{t}^2+\hat{u}^2}{\hat{s}^2}, \\
\langle \vert\mathcal{\hat{M}}(q \bar q \to q \bar q)\vert^2 \rangle
& = \frac{4}{9} \left( \frac{\hat{s}^2+\hat{u}^2}{\hat{t}^2} + \frac{\hat{t}^2+\hat{u}^2}{\hat{s}^2} \right) -\frac{8}{27}\frac{\hat{u}^2}{\hat{s}\hat{t}}, \\
\langle \vert\mathcal{\hat{M}}(q \bar q \to gg)\vert^2 \rangle
& = \frac{32}{27} \frac{\hat t^2 + \hat u^2}{\hat t \hat u} - \frac{8}{3} \frac{\hat t^2 + \hat u^2}{\hat s^2},\\
\langle \vert\mathcal{\hat{M}}(gq \to gq)\vert^2 \rangle
& = -\frac{4}{9} \frac{\hat s^2 + \hat u^2}{\hat s \hat u} + \frac{\hat s^2 + \hat u^2}{\hat t^2},\\
\langle \vert\mathcal{\hat{M}}(gg \to q \bar q)\vert^2 \rangle
& = \frac{1}{6} \frac{\hat t^2 + \hat u^2}{\hat t \hat u} - \frac{3}{8} \frac{\hat t^2 + \hat u^2}{\hat s^2},\\
\langle \vert\mathcal{\hat{M}}(gg \to gg)\vert^2 \rangle
& = \frac{9}{2} \biggl (3 - \frac{\hat t \hat u}{\hat s^2} - \frac{\hat s \hat u}{\hat t^2} - \frac{\hat s \hat t}{\hat u^2}\biggr ),
\end{align}
\end{subequations}
where we factored out the coupling constant $\vert\mathcal{M}\vert^2 = (4\pi\alpha_s(\mu_R^2))^2\vert\mathcal{\hat{M}}\vert^2$. Here, the coupling constant is evaluated at the renormalization scale $\mu_R$ (for partonic cross section we take $\mu_R=\mu_F=p_\perp$, where $p_\perp$ is the transverse parton momentum).

The single inclusive jet cross section at LO is then given as
\begin{align}
\label{eq:oneejetsigmanew}
&\frac{{d}\sigma^{AB}_{j}}{{d}y_\text{inc} p_{\perp} {d}p_{\perp}} =
\frac{1}{8\pi s^2} \sum_{a,b,c,d} \int_{y_\text{min}}^{y_\text{max}} {d}y \frac{f^A_{a}(x_A,\mu_F^2)}{x_A}\frac{f_{b}^B(x_B,\mu_F^2)}{x_B}\nonumber\\
&\times \left(4\pi \alpha_s(\mu_R^2)\right)^2 \langle \vert \mathcal{\hat{M}}(ab \to cd)\vert^2\rangle (y, y_\text{inc}),
\end{align}
where $\cosh y_\text{inc} < \frac{\sqrt{s}}{2p_\perp}$ and the integration limits for $y_\text{min} < y < y_\text{max}$ are given by,
\begin{equation}
  -\log\left( \frac{\sqrt{s} }{p_\perp}-e^{-y_\text{inc}} \right) < y <\log\left( \frac{\sqrt{s} }{p_\perp}-e^{y_\text{inc}} \right).
\end{equation}

\subsection{Single inclusive hadron spectra}

The single inclusive hadronic cross section at LO in the absence of medium modifications is given by the convolution of the jet spectrum \Eq{eq:oneejetsigmanew} with the fragmentation function $D^k_h$:
\begin{equation}
\frac{d \sigma^{AB}_{h}}{{d}y_\text{inc} \mathrm{d}p_{\perp}} = \int {d}q_\perp {d}z \frac{d\sigma^{AB}_{k}}{{d}y_\text{inc}{d}q_\perp} D_h^k(z,\mu_{F}^2)\delta(p_{\perp} - zq_{\perp}),
\end{equation}
where the cross section for producing $q_\perp$ momentum parton $k$ is convoluted with the probability to fragment to momentum $p_\perp = zq_\perp$ charged hadron.

Performing the integration over $q_{\perp}$ and inserting the partonic cross section formula, the invariant hadron spectra may be rewritten as
\begin{equation}
\begin{split}
&E_h \frac{d\sigma^{h}}{d^3p}=\sum_{c,d}
\int_{z_\text{min}}^1 \frac{dz}{z^2}\frac{1}{2}\left(D_{c}(z,\mu_{F}^2)+D_{d}(z, \mu_{F}^2 ) \right) \\
&
\times \frac{1}{16\pi^2 s^2} \sum_{a,b} \int_{y_\text{min}}^{y_\text{max}} \mathrm{d}y \frac{ f^{A}_{a}(x_A,\mu_F^2)}{x_A}\frac{f^{B}_{b}(x_B,\mu_F^2)}{x_B}\\
&
\times\left(4\pi \alpha_s(\mu_R^2)\right)^2\langle \vert \mathcal{\hat{M}}(ab \to cd)\vert^2 (y,y_\text{inc})\rangle.
\end{split}
\end{equation}
In the expression above, the momentum fractions $x_A$ and $x_B$ appearing in \Eq{eq:oneejetsigmanew} are evaluated at the rescaled momentum  $p_{\perp} \to p_{\perp}/z$ and
$
z_\text{min}=  \frac{2p_\perp}{\sqrt{s}} \cosh y_\text{inc}.
$
The gluon and the (averaged) quark fragmentation functions are given by BKK parametrization~\cite{Binnewies:1994ju} (for an implementation example see the  \texttt{INCNLO} computer code\footnote{\url{http://lapth.cnrs.fr/PHOX_FAMILY/readme_inc.html}}). That is, for simplicity,  in this paper we use a single quark FF $D_q(q) \equiv \frac{1}{2N_f}\sum_a (D_a(z)+D_{\bar{a}}(z))$ with $N_F=5$. We checked that this has only small effect on the $R_\text{AA}^h$ in \Fig{fig:RAAnpdf} for $20\,\text{GeV}<p_T<200\,\text{GeV}$.

Finally, conventionally the renormalization  and  factorization scales for hadronic spectra are taken to be $\mu_R = \mu_F = p_\perp$, where $p_\perp$ is the transverse hadron momentum. This is not identical to first calculating the partonic spectra in  \Eq{eq:oneejetsigmanew} and then convolving it with FFs due to different choice of the scale in PDFs and $\alpha_S$. We ignore this difference in the model calculations where energy loss is calculated for quarks and gluons and the resulting spectrum is convolved with  FFs.

\clearpage

\end{document}